\documentclass[floatfix,nofootinbib]{revtex4-2}

\usepackage{graphicx}
\usepackage{dcolumn}
\usepackage{bm}

\usepackage{color}
\usepackage{amsmath}
\usepackage{amsfonts}
\usepackage{amssymb}
\usepackage{latexsym}
\usepackage[center]{caption}
\usepackage{subcaption}
\usepackage{tikz}
\usepackage{circuitikz}
\usepackage[compat=1.1.0]{tikz-feynman}
\usepackage[compat=1.1.0]{tikz-feynhand}
\usepackage{hyperref}
\definecolor{darkred}{rgb}{0.8,0.1,0.1}
\hypersetup{colorlinks=true, linkcolor=darkred, citecolor=blue, linktoc=page}
\usepackage{tikz-cd,mathtools}

\def\m{\mu}
\def\n{\nu}

\def\){\right)}
\def\({\left( }
\def\]{\right] }
\def\[{\left[ }

\def\<{\langle }
\def\>{\rangle}

\newcommand{\be}{\begin{equation}}
\newcommand{\ee}{\end{equation}}

\def\bsk{{\boldsymbol{k}}}
\def\bsp{{\boldsymbol{p}}}

\def\bsw{{\boldsymbol{w}}}
\def\bsx{{\boldsymbol{x}}}
\def\bsy{{\boldsymbol{y}}}
\def\bsz{{\boldsymbol{z}}}

\def\a{\alpha}
\def\b{\beta}

\def\no{\nonumber}
\def\p{\partial}

\def\eps{\epsilon}

\tikzset{graviton/.style={decorate, decoration={snake, amplitude=.4mm, segment length=1.5mm, pre length=.5mm, post length=.5mm}, double}}

\begin{document}

\title{QCD-Gravity double-copy in the Regge regime: shock wave propagators}

\author{Himanshu Raj}
\email{himanshu.raj@stonybrook.edu}
\affiliation{
Center for Frontiers in Nuclear Science, Department of Physics and Astronomy, Stony Brook University, Stony Brook, NY 11794, USA
}

\author{Raju Venugopalan}
 \email{raju.venugopalan@gmail.com}
\affiliation{
Department of Physics, Brookhaven National Laboratory, Upton, NY 11973, USA\\
Center for Frontiers in Nuclear Science, Department of Physics and Astronomy, Stony Brook University, Stony Brook, NY 11794, USA}

\date{\today}

\begin{abstract}
\vspace{.25in}
In \cite{Raj:2023irr}, we demonstrated a double-copy relation between inclusive gluon radiation in shock wave collisions of ultrarelativistic nuclei and inclusive graviton radiation in trans-Planckian gravitational shock wave collisions. We compute here the corresponding gravitational shock wave propagators in general relativity and demonstrate that they too obey a double copy relation to gluon shock wave propagators computed previously. These results provide key input in a renormalization group approach towards computing the high frequency radiation spectrum in close black hole encounters. 
\end{abstract}

\maketitle

\section{introduction}

In two recent papers~\cite{Raj:2023irr,Raj:2023iqn}, we explored  universal features of multiparticle production in $2\rightarrow N$ gluon and graviton final states (with $N\gg 1$), respectively, in QCD and in Einstein gravity (GR). In the GR context, the program of such studies was first initiated by Lipatov~\cite{Lipatov:1982it,Lipatov:1982vv} and by Amati, Ciafaloni and Veneziano (ACV)~\cite{Amati:1987uf,Amati:1987wq,Amati:1990xe} in the language of scattering amplitudes. Our work adopts a complementary semi-classical framework of gravitational shock wave scattering.  This approach was  previously developed successfully in QCD to describe multiparticle production in hadron-hadron and electron-hadron scattering in QCD at collider energies. In particular, since the rapid growth in gluon occupancies in high energy Regge asymptotics leads to the dominance of saturated states of maximal occupancy~\cite{Gribov:1983ivg,Mueller:1985wy} in multiparticle production, their properties could be computed systematically and efficiently employing many-body methods~\cite{McLerran:1993ka,McLerran:1993ni}. This QCD framework, now called the Color Glass Condensate effective field theory (CGC EFT)~\cite{Iancu:2003xm,Gelis:2010nm,Kovchegov:2012mbw}, allows one to recover results previously obtained for $2\rightarrow N$  scattering by employing amplitude methods; a specific example is the celebrated BFKL equation of Lipatov and collaborators~\cite{Lipatov:1976zz,Balitsky:1978ic}. More generally, the CGC EFT allows one to derive powerful nonlinear renormalization group (RG) equations-the BK/JIMWLK equations ~\cite{Balitsky:1995ub,Kovchegov:1999ua,Iancu:2000hn,Iancu:2001ad,Jalilian-Marian:1997qno,Jalilian-Marian:1998tzv}- that allow for quantitative descriptions of a wide range of high occupancy final states in hadron and nuclear collisions. Such applications pose a formidable challenge for first principles Feynman amplitude computations.

In Regge asymptotics, the structure of $2\rightarrow N$ amplitudes, in both QCD and GR, can be constructed from the imaginary part of the $2\rightarrow 2$ amplitude $A(s,t)$. In the $t$-channel, this has a ladder structure. The ladder is ordered in relative rapidities (the logarithms of sub-energies of each cut rung) and is comprised of two building blocks making up, respectively, the individual rungs and legs of the ladder. The former are so-called Lipatov vertices anchoring each end of a given rung; these sum all possible $2\rightarrow 3$ amplitudes that are allowed in these asymptotics. The Lipatov vertices are nonlocal functions of the two momenta along the  legs attached to a rung. Their product, from both ends of the rung, is a gauge invariant quantity. 

The other building block is the reggeized propagator describing momentum transfer along the legs of the ladder. Reggeization in this context refers to the multi-loop iteration of the bare t-channel propagator 
$\frac{1}{ t}\rightarrow \frac{1}{t}\,e^{\alpha_R(t)\,Y}$,
where only contributions containing the leading double logarithmic contributions $\alpha_R(t)= \alpha\, \ln{t/\Lambda}$ and in the rapidity variable $Y=\ln{s/t}$ are retained. Here $\alpha$ refers to the coupling constant, $t$ and $s$ are, respectively, the squared momentum transfer and squared center-of-mass energy of a given sub-ladder, and $\Lambda$ is an infrared cutoff. The function $\alpha_R(t)$ satisfies a Regge theory interpretation as the poles $1/(j-\alpha_R(t))$ of the scattering amplitude in the $t-j$ plane, where $j$ denotes the spin of the $t$-channel exchange. Lipatov and ACV showed that these structures can be incorporated in a so-called reggeon field theory for both QCD and GR~\cite{Lipatov:1991nf}; for a recent review of this eikonal approach, see \cite{DiVecchia:2023frv}. 

In QCD, for weak coupling $\alpha \ll 1$, this BFKL construction resums leading logarithms $(\alpha Y)^N$ to all orders in perturbation theory. The intercept of this amplitude at $t=0$ corresponds to the BFKL pomeron, which describes the $2\rightarrow N$ cross-section $\sigma = 2\,{\rm Im} A(s,t=0)\propto s^{\alpha_R(t=0)}$, with $\alpha_R(t=0)\sim 0.5$, for $\alpha=0.2$. The BFKL pomeron leads to a  growth of the cross-section that is far more rapid than its nonperturbative counterpart; a smooth transition from soft to hard dynamics is clearly observed in deeply inelastic scattering (DIS) collider data~\cite{Cooper-Sarkar:2016foi}. 

An important point to note here is that the longitudinal momentum are soft in both regimes, corresponding to momentum fractions $x\ll 1$ of gluons in the $2\rightarrow N$ scattering. The soft to hard transition is in transverse momenta, where in the latter case, the gluons carry large transverse momenta $k_\perp \gg \Lambda_{\rm QCD}$. Specifically, the BFKL framework is valid for $|t|\gg \Lambda_{\rm QCD}^2$ with $|t|/s\rightarrow 0$. Systematic higher order corrections to this BFKL construction are therefore feasible. For an excellent recent summary of state-of-the art, we refer readers to \cite{DelDuca:2022skz}.

Lipatov's remarkable result in \cite{Lipatov:1982it,Lipatov:1982vv} is that the gravitational Lipatov vertex satisfies a double-copy relation 
\begin{align}
\label{grav-lipatov-vertex}
    \Gamma^{\m\n} = \frac12 C^\m C^\n -\frac12 N^\m N^\n~,
\end{align}
where $C^\mu$ is the QCD Lipatov vertex \cite{Lipatov:1976zz} and $N^\mu$ is the QED bremsstrahlung factor.  In \cite{Raj:2023irr}, we showed that this result can be recovered from a semi-classical computation of radiation emitted in the scattering of gravitational shock waves in GR. In particular, we showed that the computation is exactly analogous to prior computations of gluon radiation in the scattering of gluon shock waves in the CGC EFT, where the amplitude for gluon radiation is proportional to $C^\mu$~\cite{Blaizot:2004wu,Blaizot:2004wv,Gelis:2005pt}. Independently of the work of Lipatov, double-copy relations between Yang-Mills theory and gravity have a long history~\cite{Kawai:1985xq}, an example of this being the BCJ color-kinematic double-copy~\cite{Bern:2008qj,Bern:2019prr}; this powerful duality has recently been applied to compute high order corrections to gravitational radiation from the inspiral phase of black hole mergers~\cite{Bern:2022jvn,Damour:2019lcq}. 

While there have been interesting attempts to understand the Lipatov double-copy within the framework of the BCJ color kinematic duality~\cite{SabioVera:2012zky,SabioVera:2014mkb,Johansson:2013nsa}, we showed in \cite{Raj:2023iqn} that the Lipatov double-copy can be simply understood as a classical double-copy~\cite{Monteiro:2014cda,Goldberger:2016iau,Goldberger:2017frp,Luna:2016due,Kosower:2018adc,delaCruz:2020bbn}. Building on prior work by Goldberger and Ridgway~\cite{Goldberger:2016iau} computing gluon radiation from the scattering of classical color charges coupled to slowly varying Yang-Mills background fields~\cite{Wong:1970fu}, we observed that their classical double-copy replacement rules\footnote{As emphasized in \cite{Raj:2023iqn}, in order to recover the gravitational Lipatov vertex, it is very important to keep the sub-eikonal corrections to $C^\mu$  computed in the Yang-Mills + Wong framework. These sub-eikonal terms are crucial in correctly recovering the second term in Eq.~\eqref{grav-lipatov-vertex} proportional to the bilinear of the QED bremsstrahlung factor.} for gluon radiation from ultrarelativistic color charges $c^a$ with color index $a$, 
\begin{align}
\begin{split}
& c^a \rightarrow p^\mu \,,\\
& i f^{a_1 a_2 a_3} \rightarrow \Gamma^{\nu_1 \nu_2 \nu_3}\left(q_1, q_2, q_3\right)=-\frac{1}{2}\left(\eta^{\nu_1 \nu_3}\left(q_1-q_3\right)^{\nu_2}+\eta^{\nu_1 \nu_2}\left(q_2-q_1\right)^{\nu_3}+\eta^{\nu_2 \nu_3}\left(q_3-q_2\right)^{\nu_1}\right)~,
\end{split}
\end{align}
and $g\rightarrow \kappa$ (with $g$ the QCD coupling, and $\kappa^2= 8\pi G$, where $G$ is Newton's constant), 
leads to Eq.~\eqref{grav-lipatov-vertex}. Interestingly, formally taking the soft limit of gluon radiation in the Yang-Mills+Wong computation, where {\it both} transverse and longitudinal momenta are small, and performing the corresponding classical double-copy, leads to Weinberg's result~\cite{Weinberg:1965nx} for soft graviton radiation~\cite{PV:2019uuv}. 

Thus as emphasized by Ciafaloni et al.~\cite{Ciafaloni:2018uwe}, there are two regimes of $2\rightarrow N$ multiparticle production in gravity. One is the Weinberg regime of soft gravitons, and the other is the Lipatov regime of semi-hard ``wee" partons, in Feynman's terminology of partons with momentum fractions $x\rightarrow 0$ but transverse momenta $\sim 1/R_S$, where $R_S$ is the Schwarzchild radius. As we discussed at length in 
\cite{Raj:2023irr}, inelastic production of semi-hard gravitons becomes significant for impact parameters $b$ satisfying $\sqrt{|t|}\sim 1/b\leq 1/R_S$.

In QCD,  the copious growth in gluon occupancy generates many-body screening and recombination effects that regulate the growth in the occupancy to saturate at a maximal occupancy $N\sim 1/\alpha$. The overoccupied gluon fields form a semi-classical lump characterized by an emergent close packing semi-hard scale $Q_S(x)$. It is their many-body dynamics that is captured by the BK/JIMWLK equations, with $Q_S(x)$ the scale determined by the nontrivial fixed point of their RG evolution with rapidity. The classical lump also saturates unitarity perturbatively at scales $1/Q_S \ll 1/\Lambda_{\rm QCD}$. In ultrarelativistic heavy-ion collisions, it is the formation, decay and rescattering of gluons forming these classical lumps that leads to the formation of the quark-gluon plasma~\cite{Berges:2020fwq}.

The outstanding question of interest to us is whether an analogous picture emerges in $2\rightarrow N$ scattering in gravity for impact parameters $b\rightarrow R_S$. Indeed, it has been argued previously that black holes can be understood as overoccupied self-bound semi-classical graviton lumps, with the Schwarzchild radius representing the close packing scale $R_S\sim 1/Q_S$~\cite{Dvali:2010jz,Dvali:2011aa}. Simple power counting in this Black Hole Quantum Portait (BHQP) picture suggests that the probability to form such semi-classical lumps in $2\rightarrow N$ processes is exponentially suppressed unless the corresponding microstate entropy is $O(1/\alpha)$; this has been argued to saturate the Bekenstein bound in units of the Goldstone decay scale of these microstates~\cite{Dvali:2020wqi}. As noted in \cite{Dvali:2021ooc}, this reasoning is remarkably similar to the understanding of the emergence of semi-classical gluon lumps in the CGC EFT. It was argued further that the underlying physics of such semi-classical lumps that saturate unitarity is universal and independent of the underlying dynamics. Specifically, the CGC-BHQP correspondence suggests an analogous Goldstone picture of modes in the CGC EFT representing the overoccupied gluon microstates. Some of the consequences in the CGC EFT context have been noted separately~\cite{Raj:2023zxo,Nair}. 

One way to confirm this conjecture of universality across widely separated energy scales  is to develop a similar quantitative RG analysis for gravitational radiation in the Lipatov regime, as was previously developed for gluon radiation in the CGC EFT. In  concrete terms, can we\\
I) derive the gravitational equivalent of the BFKL equation in the shock wave scattering formalism,\\
II) generalize this study to derive the equivalent of the BK/JIMWLK equations in gravity, and \\
III) observe the emergence of a nontrivial fixed points corresponding to black hole formation.\\
Questions I and II are of phenomenological interest in the computation of  gravitational radiation from the merger phase of black hole scattering where velocities reach ultrarelativistic speeds and numerical relativity simulations become extremely challenging~\cite{Page:2022bem}. Developing quantitative methods to address the final instants of gravitational radiation~\cite{Pretorius:2007jn,Sperhake:2012me} is an important practical motivation for this work. 

We observed earlier that the key ingredients in question I
are the Lipatov vertices and the reggeized propagators. In \cite{Raj:2023iqn,Raj:2023irr}, we computed the former in the shock wave formalism. The focus of this work will be the computation of  shock wave propagators; the mapping of these to the language of graviton propagators has been discussed previously~\cite{Hentschinski:2018rrf,Bondarenko:2017ern,Bondarenko:2020kwm}. In subsequent work, we will employ these propagators to develop the RG framework necessary to address questions I through III. Our discussion throughout will be grounded in the Schwinger-Keldysh formalism of cut and retarded propagators in quantum field theory rather than in in the language of Feynman amplitudes alone or in that of regggeon field theory, with the latter, as noted, employed in the Lipatov-ACV approach~\cite{Lipatov:1991nf,Lipatov:1996ts,Amati:1992zb,Amati:1993tb,Amati:2007ak}. As observed in \cite{Gelis:2006cr,Gelis:2006yv}, the so-called Abramovsky-Gribov-Kancheli (AGK) cutting rules~\cite{Abramovsky:1973fm} employed extensively by ACV can be understood in the Schwinger-Keldysh formalism (SK) as a specialized application of Cutkosky's rules in the presence of time dependent sources. Further, as emphasized in an elegant discussion in \cite{Caron-Huot:2010fvq}, such SK methods are closely related to the generalized unitarity based approaches to scattering amplitude computations~\cite{Bern:1994cg,Britto:2005fq}.

The SK methods we will adopt are especially suited to the computation of gravitational radiation. In particular, we will be interested in the computation of the inclusive radiation spectrum as a function of impact parameter and rapidity. It was shown in \cite{Gelis:2006cr,Gelis:2006yv} that the computation of inclusive distributions can be formulated systematically in powers of the coupling $\alpha$ as an initial value problem where the ingredients are retarded and cut propagators. For gravitational (and gluon) radiation in strong backgrounds, the leading order spectrum is governed by the Lipatov vertex which, as we observed, can be computed from the equations of motion in classical shock wave scattering. To go beyond, and include the rapidity evolution of this spectrum, as well as rescattering contributions that become important as $b\rightarrow R_S$, we will need to compute next-to-leading order (NLO) corrections\footnote{As we will discuss in section IV, these NLO contributions in the shock wave formalism are one order beyond the 2-loop H-diagram of ACV in perturbation theory.} in the coupling $\alpha$. This is where the computation of shock wave propagators is essential. As we will discuss, in order to obtain these propagators, one needs to solve the initial value problem of computing the solutions of the classical GR shock wave equations of motion as well as  solutions of small fluctuation equations. 

The paper is organized as follows. In section II, we will review the derivation of results in the CGC EFT for colored scalar and gauge field propagators in the background of a gluon shock wave. We shall focus our derivation on expressions for propagators in the eikonal limit. In section III, we will present results for the full off-shell propagator of scalar and the graviton fields at the leading eikonal order. We restrict our derivation here primarily to eikonal shock wave propagators. This limit will likely be sufficient as long as one is computing radiative corrections that are not too close to the event horizon. We will also comment on the sub-eikonal corrections which are substantially more involved; these will be important in addressing question III articulated above. 
In section IV, we will sketch the computation of the NLO corrections in GR following guidance from previous such computations in the QCD.  Detailed computations carrying out the program outlined earlier will be undertaken in subsequent work. 
Appendix A provides a brief introduction to Schwinger-Keldysh propagators. The convolution properties of Green functions are  outlined in appendix B.

\section{Shock wave propagators for QCD in Regge asymptotics}
\label{sec:QCD}

In this section, we will first review the construction of eikonal shock wave propagators in QCD in high energy (Regge) asympotics. As noted in the introduction, these quark and gluon propagators are a key feature of the derivation of the BK/JIMWLK RG equations. Shock wave propagators in QCD were computed in several papers \cite{McLerran:1994vd, Balitsky:1995ub, Ayala:1995hx, Ayala:1995kg, Hebecker:1998kv,McLerran:1998nk, Balitsky:2001mr,Blaizot:2004wu, Blaizot:2004wv} using various techniques in both position and momentum space. In what follows, we shall review the derivation of a scalar propagator in the background of the CGC shock wave closely following the discussion in \cite{Blaizot:2004wu, Blaizot:2004wv}. The result for the gluon propagator analogously follows that we will present at the end of the section. This will be relevant for the classical double-copy map of this propagator to the corresponding propagator in gravity. 

In Lorenz gauge $\partial_\mu A^\mu=0$, the shock wave background of overoccupied gluons appearing in the CGC effective field theory takes the form \cite{Raj:2023irr}:
\begin{align}
\label{gauge-background}
    A_-(x^-, \bsx) = -g \delta(x^-)\frac{\rho(\bsx; y)}{\square_\perp}~.
\end{align}
Here $\rho(\bsx; y)$ denotes the color charge density generated by fast quarks and gluons carry large longitudinal momentum fractions $x\sim 1$ of the hadron momentum; these couple coherently to wee (small $x\ll 1$) gluons, corresponding to a rapidity scale $y$. This color charge density is classical due to the large occupancy, and parametrically of order $O(1/g)$, where $g$ is the QCD coupling. The background it couples to is a singular solution of the classical Yang-Mills (YM) equations in the presence of the static light cone current corresponding to the color charge density. 

The solution of the YM equations also make clear that the field strengths of the shock wave vanish outside a narrow region in $x^-$ outside $x^-=0$. The corresponding pure gauge field configurations are not identical in the spacetime region separated by the shock wave since one cannot perform a continuous gauge transformation between the two sides of the shock wave. This is manifest in light cone gauge $A^+=0$, obtained by performing the gauge transformation $A_\m \to U A_\m U^\dagger +\frac{i}{g} U\p_\m U^\dagger$ on the solution in Eq.~\eqref{gauge-background}.

In light cone gauge,  only transverse components of the gauge field are nonvanishing and are given by
\be
\label{gauge-background-theta}
A_i =\Theta(x^-)\, \frac{i}{g}  \tilde{U}\p_i \tilde{U}^\dagger \qquad \text{where} \qquad \tilde{U} = \exp\(-ig^2 \frac{\rho(\bsx;y)}{\square_\perp}\)~.
\ee
This expression makes transparent that the vacua are not identical in the regions $x^->0$ and $x^-<0$ and correspond to distinct pure gauge solutions that cannot be transformed into one another\footnote{This property of shock wave solutions in gauge theories (and gravity) has deep connections to soft theorems and memory effects~\cite{Strominger:2017zoo}, as we will discuss further later.}. 

The direct computation of shock wave propagators for the light cone gauge $A^+=0$ (where $P^+$ corresponds to the direction of large momentum) is nontrivial, as shown in \cite{Ayala:1995kg}. However they are considerably simpler in the ``wrong" light cone gauge $A^-=0$~\cite{McLerran:1994vd}, which shares the same background field as the Lorenz gauge solution in  Eq.~\eqref{gauge-background-theta}. This background field also has a simple double-copy relation to gravitational shock waves~\cite{Akhoury:2013yua}; we will therefore henceforth work in this gauge.

We will first compute the (retarded) propagator of an adjoint scalar field coupled minimally to this background. A necessary ingredient is the solution to the small fluctuation equations of the scalar field, given by\footnote{The scalar equations of motion are
\begin{align}
    D_\mu D^\mu \phi = 0~,\qquad D_\m\phi = \p_\m\phi - ig [A_\m, \phi]~.
\end{align}
}
\begin{align}
    \square \phi - 2ig[A_-,\p_+\phi]=0~,
\end{align}
where $\square = \p_\mu \p^\mu$ is the free Laplacian. Using $[A, B]_a=i f^{a b c} A_b B_c=-\left(T^b\right)_{c a} A_b B_c=[(A \cdot T) B]_a$, we can rewrite the commutator in the above equation,
\begin{align}
\label{scalar-gauge-small-fluct}
    \square \phi(x)-2ig(A_-(x)\cdot T) \partial_{+}\phi(x)  = 0~.
\end{align}
Hence the Green function satisfies (the subscript $R$ stands for retarded)
\begin{align}
\label{retarded-prop-conv}
    \square_x G_R(x,y) - 2ig(A_-(x)\cdot T) \partial_{x^+} G_R(x,y) = \delta(x-y)~,
\end{align}
supplemented by boundary data specified at a fixed $x^-$ slice in the region $x^-<0$. As shown in \cite{Blaizot:2004wu} (and  reviewed in appendix \ref{sec:AppB}), $G_R(x,y)$ satisfies the recursive equation,
\begin{align}
\label{ret-gauge-prop-rec}
    G_R(x, y)=\int d^4 z \,G_R(x, z)\, \delta (z^- -z^-_0)\,2\,\partial_z^{-} G_R(z, y)~.
\end{align}
The surface $z^- = z^-_0$ specifies the spacetime slice where the initial data is defined.

Since the shock wave background in Eq.~\eqref{gauge-background} is pure gauge everywhere, except at $x^-=0$, the propagators $G_R(x, y)$ in this background too are free propagators everywhere, except for paths corresponding to $x^->0, y^-<0$ or its converse. For these configurations, the propagator encounters the shock wave background, and is therefore nontrivial. A  straightforward way to calculate $G_R(x, y)$ for $x^->0, y^-<0$ is to use Eq.~\eqref{ret-gauge-prop-rec} to write it as 
\begin{align}
\label{G-conv-0}
G_R(x,y) &=\int d^4z ~d^4w \,G_R^0(x,z)\, \delta(z^{-}-z^-_0)\,2\,\p^-_z\, G_R(z,w)\, \delta(w^{-}-w^-_0)\,2\,\p^-_w\,G_R^0(w,y)~,
\end{align}
and set $z_0^-=\delta$, $w^-_0=0$, where $\delta$ specifies the shock wave width in the $x^-$ direction, a parameter we will set to zero at the end of the calculation. This choice of $z^-_0$ and $w^-_0$ allows us to  replace  $G_R(x,z)$ and $G_R(w,y)$ by their corresponding free expressions $G_R^0(x,z)$, $G_R^0(w,y)$ in Eq.~\eqref{G-conv-0}, that are given by
\be
\label{free-retarded-scalar-propagator}
G_R^0(x,y) = -\int \frac{d^4k}{\(2\pi\)^4} \frac{e^{-ik\cdot(x-y)}}{k^2+i\epsilon k^-} = \frac{1}{2\pi}\Theta(x^--y^-)\Theta(x^+-y^+)\delta\((x-y)^2\)~.
\ee
The intermediate steps in obtaining the last equality are provided in appendix \ref{app:B2}. 

In order to compute $G_R(z,w)$ in Eq.~\eqref{G-conv-0}, we begin with the definition of the retarded Green function,
\be
\label{retarded-gf-0}
G_R(x, y) = -\int \frac{d^4k}{\(2\pi\)^4} \frac{1}{k^2+i\epsilon k^-} \phi_k(x)\phi_k^*(y)~,
\ee
where $\phi_k(x)$ is the eigenfunction of the small fluctuation equation in Eq.~\eqref{scalar-gauge-small-fluct}. In the absence of the shock wave background, the eigenfunctions are simply plane waves, and we recover  Eq.~\eqref{free-retarded-scalar-propagator}. 

To solve for $\phi_k(x)$ in the neighborhood of the shock wave (in the $x^-$ direction), we make the eikonal assumption that changes in the field in the transverse directions are negligible compared to those in the longitudinal direction. We can then approximate Eq.~\eqref{scalar-gauge-small-fluct} to read 
\be
\label{scalar-gauge-small-fluct-1}
2\p_+\p_-\phi(x) -2ig(A_-(x) \cdot T) \partial_{+}\phi(x) = 0\,.
\ee
With the input from the scalar field solution in  the region $x^-<0$, the solution for $x^- > 0$ is 
\be
\phi^a(x^->0) = U(\bsx)^a_{~b} ~\phi^b(x^-<0)~,\qquad {\rm with}\qquad \phi^a(x^-<0) =  e^{-ikx} c^a~.
\ee
Here $c^a$ is a pointlike color charge and $U(\bsx)$ is a light-like Wilson line, path ordered in $x^-$: 
\be
\label{eq:QCD-Wilsonline}
U(\bsx) = P \exp \(ig\int dz^- A_-(z^-, \bsx)\cdot T\)~.
\ee
Thus, in the eikonal approximation, the result of the scattering of the  colored scalar with the shock wave is to simply rotate it in color space. 
 
In summary, the solution of the small fluctuation scalar equation in the eikonal approximation is 
\be
\phi_k^a(x) = \Theta(-x^-) \,e^{-ikx} c^a + \Theta(x^-)\, e^{-ikx} U^a_{~b}(\bsx) c^b~.
\ee
Inserting this result into Eq.~\eqref{retarded-gf-0} gives the free propagator for $\theta(-x^-)\theta(-y^-)$ or the color rotated free propagator for $\theta(x^-)\theta(y^-)$. For the  nontrivial component of the Green function in the r.h.s of Eq.~\eqref{G-conv-0}, the term crossing the shock wave is given by 
\begin{align}
\label{scalar-non-trivial}
    G_R^{ab}(z^->0, ~w^-<0) = -\int \frac{d^4k}{\(2\pi\)^4} \frac{e^{-i k\cdot(z-w)}}{k^2+i\epsilon k^-} c^a \[U(\bsz)\]^b_{~d} c^d~.
\end{align}
Note that in Eq.~\eqref{G-conv-0} we need to take the limit $\delta\to 0$. This gives 
\begin{align}
    \lim_{\delta\to0^+} G_R^{ab}(z^-=\delta, ~w^-=0) &= -\lim_{\delta\to0^+} \int \frac{d^4k}{\(2\pi\)^4} \frac{e^{-i k^+\delta-i k^-(z^+-w^+)+i \bsk\cdot(\bsz-\bsw)}}{k^2+i\epsilon k^-} c^a \[U(\bsz)\]^b_{~d} c^d~,\no\\
    &= -\lim_{\delta\to 0^+} \int \frac{d^4k}{\(2\pi\)^4} \frac{e^{-i k^+\delta-i k^-(z^+-w^+)+i \bsk\cdot(\bsz-\bsw)}}{2k^-\(k^+-\frac{\bsk^2}{2k^-}+i\frac{\eps}{2}\)} c^a \[U(\bsz)\]^b_{~d} c^d~,\no\\
    &= \frac12 \Theta(z^+-w^+)\delta^{(2)}(\bsz-\bsw) c^a \[U(\bsz)\]^b_{~d} c^d~.
    \label{sw-prop}
\end{align}
To obtain this result, we first performed the $k^+$ integral by closing the contour from below, and then took the limit $\delta\to 0^+$. We also used the identity
\be
\int \frac{dk^-}{2\pi} \frac{e^{-ik^-(z^+-w^+)}}{-ik^-} = \Theta(z^+-w^+)~.
\ee
Plugging this result into Eq.~\eqref{G-conv-0}, we obtain 
\begin{align}
\label{scalar-gauge-prop}
G_R(x,y) &=\int d^4z ~G_R^0(x,z)\, \delta(z^{-})\, U(\bsz)\, 2\,\p^-_z G_R^0(z,y)~,
\end{align}
which agrees\footnote{\label{ret-Feyn}In \cite{Hebecker:1998kv}, the authors are computing the Feynman propagator instead of the retarded propagator; this explains the $U^\dagger$ in their Eq.~(9) instead of $U$ in our case. See  \cite{Blaizot:2004wv} for a discussion retarded versus Feynman propagators in this context.} with Eq. (9) of  \cite{Hebecker:1998kv}. 

The Fourier transform 
\begin{align}
\label{fourier-transform-def}
    \tilde{G}_R(p,p') = \int d^4x~d^4y~ G_R(x,y) e^{ipx}e^{-ip'y}\,,
\end{align}
of the propagator then takes the compact form 
(derived in appendix \ref{sec:AppB})
\begin{align}
\label{scalar-gauge-prop-mom}
    \tilde{G}_R(p,p') = \tilde{G}_R^0(p) (2\pi)^4\delta^{(4)}(p-p') + \tilde{G}_R^0(p) \mathcal{T}(p,p')\tilde{G}_R^0(p')\,,
\end{align}
where $\tilde{G}_R^0(p)$ is the free propagator in momentum space,
\be
\label{free-G0-prop}
\tilde{G}_R^0(p) = \int d^4x ~e^{ipx} ~G^0_R(x) = -\frac{1}{p^2+i\epsilon p^-}~,
\ee
and the effective vertex is 
\be
\label{T-matrix-g}
\mathcal{T}(p,p') \equiv - 4\pi i(p')^-\delta(p^- -(p')^-)  \(\int d^2\bsz ~e^{i(\bsp-\bsp')\cdot \bsz}\(U(\bsz)-1\) \)~.
\ee
The dressed propagator can be represented pictorially as 
\begin{figure}[ht]
\centering
\raisebox{-12pt}{\includegraphics[scale=1.0]{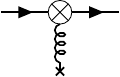}}
=
\raisebox{-0pt}{\includegraphics[scale=1.0]{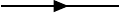}}
+ $\sum_{n=1}^\infty$
\raisebox{-12pt}{\includegraphics[scale=1.0]{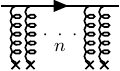}}
\caption{Dressed propagator of a scalar field in the shock wave background of overoccupied gluons. The first term on the r.h.s is the free scalar propagator corresponding to the no scattering ``1" term in the expansion of the Wilson line in Eq.~\eqref{T-matrix-g}. The other terms in the expansion of the Wilson line are the contributions from coherent multiple scatterings of the scalar field off color charges within the shock wave.}
\label{gluon-shock wavepropagator}
\end{figure}

The results  derived above for  the colored scalar propagator in the shock wave background can be straightforwardly extended to fermion and gluon propagators in the gluon shock wave background; these were first computed in \cite{McLerran:1994vd,Balitsky:1995ub,McLerran:1998nk} in position space and later given in momentum space~\cite{Balitsky:2001mr}. 
Indeed, the gluon shock wave propagator in the lightcone gauge $A_+=0$ can be written similarly to Eq.~\eqref{scalar-gauge-prop-mom} as 
\begin{align}
\label{gluonPropPosFinal2}
    \tilde{G}_{R,\mu \nu}\left(p, p^{\prime}\right)=(2 \pi)^4 \delta^{(4)}\left(p-p^{\prime}\right)\tilde{G}^0_{R,\mu \nu}(p)+\tilde{G}^0_{R,\mu \rho}(p) \mathcal{T}_R^{\rho \sigma}\left(p, p^{\prime}\right)\tilde{G}^0_{R,\sigma \nu}\left(p^{\prime}\right)~,
\end{align}
where the effective vertex is 
\cite{Balitsky:2001mr, Roy:2018jxq, Roy:2019hwr, Blaizot:2004wv}
\begin{align}
\label{T-matrix-gluon}
    \mathcal{T}_R^{\mu \nu}\left(p, p^{\prime}\right)=-4 \pi i \Lambda^{\mu \nu} p^- \delta\left(p^{-}-(p')^{-}\right) \int d^2 \boldsymbol{z}~ \mathrm{e}^{i \boldsymbol{z} \cdot\left(\boldsymbol{p}-\boldsymbol{p}^{\prime}\right)}\left( U\left(\boldsymbol{z}\right)-1\right)~.
\end{align}
Here $\tilde{G}^0_{R,\mu \nu}(p) = \Lambda_{\m\n}\tilde{G}^0_{R}(p)$ with $\Lambda_{\m\n} = \eta_{\m\n}-\frac{p^\m n^\n+p^\n n^\m}{p\cdot n}$. 
We emphasize that for phenomenological computations this light cone gauge $A_+=0$ should be distinguished from the light cone gauge $A_-=0$, where the parton interpretation of structure functions extracted from experiments (with hadron momenta $P^+$) is more transparent. 

The result in Eq.~\eqref{gluonPropPosFinal2} follows in a similar manner to the scalar case. One defines the spin-1 shock wave propagator in terms of the small fluctuation fields as 
\begin{align}
\label{gaugePropDef1}
    G_{\m\n}(x, y)=  -\int \frac{d^4k}{\(2\pi\)^4} \frac{1}{k^2+i\epsilon k^-}\sum_{\lambda} a_{\m, k}^{(\lambda)}(x)a_{\n, k}^{*(\lambda)}(y)~.
\end{align}
After gauge fixing to light cone gauge $a_+=0$, the solution to the small fluctuation field $a_{\m, k}^{(\lambda)}(x)$ is readily determined~\cite{Gelis:2005pt}. 
Asymptotically, the solutions are plane waves $a_{\m, k}^{(\lambda)}(x^-<0) = e^{-ikx} \,\epsilon_{\m}^{(\lambda)}(k)$, where $\epsilon_{\m}^{(\lambda)}(k)$ is the polarization tensor in light cone gauge. In arriving at the result for the gluon shock wave propagator, one employs the completeness relation 
\begin{align}
\label{gauge-pol-completenessrel}
    \sum_{\lambda} \epsilon_{\m}^{(\lambda)}(k) \(\epsilon_{\n}^{(\lambda)}(k)\)^* = \eta_{\m\n} -\frac{k_\m n_\n+ k_\n n_\m}{k\cdot n}~,
\end{align}
where $n$ is a null four vector such that $\epsilon\cdot n = 0$ and $k\cdot n\neq 0$. Note that this completeness relation is identical to the one in the absence of shock wave background and will not hold for arbitrary background gauge fields. The reason we can use it here is because in the eikonal approximation the interactions of small fluctuation fields with the shock wave do not modify the polarization tensor at leading eikonal order. A similar reasoning will be used in the gravity discussion in the next section.

\section{Shock wave propagators for gravity in Regge asymptotics}
\label{sec:gravity}

We will now follow the methods outlined in the previous section on QCD at high energies and apply these to construct shock wave propagators in gravity. After first reviewing the properties of gravitational shock waves, we will compute scalar propagators in shock wave spacetime and follow this by a computation of the graviton shock wave propagator. The latter computation will be performed in the eikonal regime where graviton polarizations are not affected by interactions with the shock wave. We will comment on sub-eikonal modifications to these results. We will also briefly comment on classical double-copy relations to the Yang-Mills results.


\subsection{Gravitational shock waves}
\label{sec:2.1}
When a Schwarzschild black hole characterized by mass $m$ is given an infinite boost $\gamma \to \infty$, then in the limit $m\to 0$, with total energy $\mu= \gamma\, m $ held fixed, one obtains the shock wave spacetime \cite{Aichelburg:1970dh}
\begin{align}
\label{ASmetric}
ds^2 = &~ 2\,dx^+dx^- -\delta_{ij}dx^i dx^j + 8\,\mu \,G_N \,\delta(x^-)\log(\Lambda |\bsx|) \(dx^-\)^2~.
\end{align}
Here $G_N$ is the Newton coupling and $\Lambda$ is an IR scale. This metric is a solution to Einstein's equation with a nonvanishing energy-momentum tensor given by $T_{\m\n} = \delta_{\mu-}\delta_{\nu-} \delta(x^-) \delta^{(2)}(\bsx)$, which corresponds to a massless point particle located at the origin of the transverse space $\bsx=0$. This  point-like distribution in transverse space implicitly assumes a distance scale over which its internal distribution is not resolved. However in general, since GR is nonlinear, a highly energetic particle will carry a cloud of gravitons corresponding to a nontrivial transverse density distribution $\rho(\bsx)$. 

A systematic field theory treatment, analogous to that outlined for the gauge theory case in the previous section, would therefore entail starting with the point particle description which then evolves in rapidity from the momentum $P^+$ (corresponding to the support of the delta function in $x^-$) to a lower momentum scale $\Lambda^+$ through the emission of slower gravitons with $\Lambda^+< k^+ < P^+$. As with the case of the classical color charge density, one can similarly assume that $\rho(\bsx)$ is specified at this scale, which can then be evolved systematically with decreasing $\Lambda^+$ to accommodate the growth in $\rho(\bsx)$ with further emissions of slower gravitons. At a given $\Lambda^+$, one can generalize the form of the EM tensor to read 
\begin{align}
\label{EMtensor1}
T_{\m\n} = \delta_{\mu-}\delta_{\nu-}\, \mu \,\delta(x^-)\, \rho(\bsx)~,
\end{align}
with the support of the $\delta$-function is understood to be $1/\Lambda^+$. 
The resulting spacetime has the metric
\be
\label{denseBgnd1}
ds^2 = 2dx^+dx^- -\delta_{ij}dx^i dx^j + g_{--}(x^-,\bsx)\(dx^-\)^2 ~,
\ee
where the nontrivial shock wave information is contained in  $g_{--}\left(x^{-}, \boldsymbol{x}\right)$, defined to be  
\be
 g_{--} \left(x^{-}, \boldsymbol{x}\right) = 2 \kappa^2 \mu \,\delta\left(x^{-}\right) \frac{\rho(\boldsymbol{x})}{\square_{\perp}}=\frac{\kappa^2}{\pi} \mu\, \delta\left(x^{-}\right) \int d^2 \boldsymbol{y}~ \ln \(\Lambda|\boldsymbol{x}-\boldsymbol{y}| \)\rho(\boldsymbol{y})~,
\ee
where $\kappa^2=8\pi G_N$. In the second equality, we reexpressed the two-dimensional Laplacian $\square_\perp\equiv\delta_{ij}\partial_i\partial_j$, in terms of the corresponding Green function.  

In this singular form, where the source $\rho$ appears linearly and a delta function arises in the metric, the spacetime remains flat both in front of ($x^-<0$) and behind ($x^->0$) the shock wave.  The inertial frames in these regions are however not identical though they are connected through a discontinuous coordinate transformation of the Minkowski vacuum. This is identical to our discussion in section~\ref{sec:QCD} of the discontinuity of large gauge transformations (in light cone gauge) across the gluon shock wave.
In GR, this is especially intuitive, given that the passing shock wave should impact spacetime measurements differently in these regions \cite{Dray:1984ha,tHooft:1996rdg}.

To rigorously analyze this, we change to the $y$-coordinate frame which is linked to the $x$-coordinate via the discontinuous transformation
\begin{align}
\label{xytransformation}
\begin{split}
x^- =& ~ y^-,\qquad x^i = ~ y^i - \kappa^2\m \,y^- \Theta(y^-)\frac{\p_i}{\square_\perp} \rho(\bsy) ~,\\[5pt]
x^+ =& ~ y^+ - \kappa^2\,\m\, \Theta(y^-) \frac{\rho(\bsy)}{\square_\perp} + \frac12  \kappa^4 \m^2~y^-\Theta(y^-) \(\frac{\p_i}{\square_\perp} \rho(\bsy)\)^2~.
\end{split}
\end{align}
These expressions comes from analyzing null-geodesics in the gravitational shock wave background. The metric in this coordinate system takes the form
\begin{align}
\label{denseBgnd2}
    ds^2 = 2dy^+dy^- -g_{ij}dy^idy^j~,
\end{align}
where $g_{ij}$ has a nonlinear dependence on the source $\rho(\bsx)$ and is given by
\begin{align}
    g_{ij} = &~ \delta_{ij}-y^-\Theta(y^-)\bigg[2\kappa^2\m~\frac{\p_i\p_j}{\square_\perp}\rho(\bsy)- \kappa^4 \m^2~y^-\(\frac{\p_i\p_k}{\square_\perp}\rho(\bsy)\) \(\frac{\p_j\p_k }{\square_\perp}\rho(\bsy)\)  \bigg]~.
\end{align}
The continuous form of the metric in Eq.~\eqref{denseBgnd2} clearly demonstrates that the region preceding the shock ($y^-<0$) corresponds to the Minkowski vacuum, while the region following the shock ($y^->0$) is a pure gauge transformation of the Minkowski vacuum (the Riemann tensor $R_{\m\n\rho\sigma}$ of the metric in the $y^->0$ region vanishes despite nonvanishing connection coefficients). As suggested earlier, this bears  exact analogy to the CGC shock wave solution described in the previous section where the field strength tensor vanishes both before and after the shock wave, even though the gauge fields are nonvanishing representing distinct pure gauge solutions separated by the gluon shock wave~\cite{McLerran:1993ka,McLerran:1993ni}.

 In a recent paper \cite{He:2023qha}, the authors established a connection between the time delay experienced by a particle propagating through a shock wave background and the gravitational memory effect. They showed that the memory mode, or the leading soft graviton mode, is related to the shock wave momentum introduced by 't Hooft in the context of scattering near a black hole horizon. Furthermore, they identified the canonically conjugate operator postulated by 't Hooft to correspond to the location of the black hole horizon with a Goldstone mode. The authors explored the implications of this relation, suggesting that it may lead to constraints on the memory effect and a connection between shocks and coherent clouds of gravitons. In our case, we see that such a description holds not just for soft gravitons {\it a la} Weinberg~\cite{Weinberg:1965nx} but also for wee gravitons {\it a la}
 Lipatov. We note further that not just ultrasoft gluons at null infinity satisfy a color memory effect~\cite{Pate:2017vwa}, but so too do wee gluons, when treated as effectively static modes on high energy scattering time scales~\cite{Ball:2018prg}. The consequences of the latter memory-like effect may have observational consequences in collider environments~\cite{Raj:2023zxo,Nair,Li:2024fdb}, and is a clearly a topic that demands further attention.


\subsection{Scalar shock wave propagators}

We will now compute the propagator for scalar fields in this spacetime. We begin with the definition from section~\ref{sec:QCD},
\begin{align}
\label{propDef1}
    G_R(x, y) = -\int \frac{d^4k}{\(2\pi\)^4} \frac{1}{k^2+i\epsilon k^-} \phi_k(x)\phi_k^*(y)~,
\end{align}
where the eigenfunctions $\{\phi_k(x)\}$ form a complete set of states. Asymptotically, these are free wave packets $\phi_k(x) = e^{-ikx}$. This equation satisfies $\square_x G(x, x') = \delta^{(4)}(x-x)$. 
As discussed in section~\ref{sec:QCD} and in appendix \ref{sec:AppB}, the retarded Green function of a scalar field (in an arbitrary background) satisfies the general property
\begin{align}
\begin{split}
G_R(x,y) &= \int d^4z~ d^4w ~G_R(x,z) \,\delta(z^{-}-z_0^-)\,2\,\p^-_z G_R(z,w) \delta(w^{-}-w^{-}_0)\,2\,\p^-_w \,G_R(w,y)~.
\end{split}
\end{align}
Adapting this expression to shock wave spacetime, we choose as previously, $z^-_0=\delta$ and $w^-_0=0$, where $\delta>0$ is introduced to give an infinitesimal width to the gravitational shock wave in the $x^-$ direction;  we will eventually set it to zero. Because we are interested in propagation through the shock wave, we take $x^->\delta$ and $y^-<0$. Thus in the regions $-\infty <y^-<0$ and $\delta <x^-<\infty$, we can replace the propagators $G_R$ with the free propagator $G^0_R$
\begin{align}
\label{G-conv-1}
G_R(x,y) &=\int d^4z ~d^4w ~G_R^0(x,z)\, \delta(z^{-}-\delta)\,2\,\p^-_z \,G_R(z,w) \delta(w^{-})2\p^-_w \,G_R^0(w,y)~.
\end{align}
The only term that we need to determine therefore is $G_R(z,w)$, for which use the definition in Eq.~\eqref{propDef1}. We will now follow an identical path to that described in section~\ref{sec:QCD}. First, we need to solve the small scalar fluctuation equations in the background of the gravitational shock wave. This was already performed in \cite{Raj:2023irr} with the result,
\begin{align}
    \label{scalar-full-solution}
    \phi_k(x) = \Theta(- x^-)\, e^{-ikx} + \Theta(x^-)\, e^{-ikx}  \,U_k(\bsx) ~,
\end{align}
where the ``gravitational Wilson line" $U_k(\bsx)$ is~\cite{Melville:2013qca,White:2017mwc} 
\begin{align}
\label{grav-WL}
    U_k(\bsx) = \exp \bigg(i\kappa^2 \mu  \frac{\rho(\bsx)}{\square_{\perp}}k^- \bigg)~.
\end{align}
As emphasized already in \cite{Raj:2023irr}, this expression is a double-copy of Eq.~\eqref{eq:QCD-Wilsonline}, with $T^a\rightarrow k^-$ and $g\rightarrow \kappa$.
Plugging this result into the definition of Green's function with $x^->\delta$ and $y^-<0$ gives
\be
\label{non-trivial-prop}
G_R(x,y) = -\int \frac{d^4k}{\(2\pi\)^4} \frac{e^{-ik(x-y)}}{k^2 +ik^-\epsilon} U_k(\bsx)~.
\ee
To compute Eq.~\eqref{G-conv-1}, we just need to evaluate Eq.~\eqref{non-trivial-prop} in the limit
\begin{align}
\label{temp1}
\begin{split}
\lim_{\delta\to 0 }G_R(x^-=\delta,y^-=0) &= -\lim_{\delta\to 0^+ }\int \frac{d^4k}{\(2\pi\)^4} \frac{e^{-i(k^-(x^+-y^+) + k^+\delta - \bsk\cdot(\bsx-\bsy))}}{k^2 +ik^-\epsilon} U_k(\bsx)~,\\
&=-\int \frac{dk^- d^2\bsk}{\(2\pi\)^3}\frac{i}{2k^-}e^{-i(k^-(x^+-y^+)  - \bsk\cdot(\bsx-\bsy))}U_k(\bsx)~,\\
&= \frac12 \Theta(x^+-y^+ - f_1) \delta^{(2)}(\bsx-\bsy) = \frac12 \,e^{f_1\p_{y^+}}\, \Theta(x^+-y^+) \,\delta^{(2)}(\bsx-\bsy)~,
\end{split}
\end{align}
where we introduced the shorthand notation,
\be
f_1(\bsx) = \kappa^2 \mu  \frac{\rho(\bsx)}{\square_{\perp}}~.
\ee
Plugging the result in Eq.~\eqref{temp1} into Eq.~\eqref{G-conv-1}, and performing the $w$-integral we obtain 
\begin{align}
    G_R(x,y) = G_R^0(x,y) + \int d^4z~ G_R^0(x,z) \(e^{-f_1(\bsz)\p_{z^+}}-1\)\delta(z^-) 2\p_{z^+} G_R^0(z,y)~.
\end{align}
Here we separated out the identity term from $e^{-f_1(\bsz)\p_{z^+}}$. Finally, upon Fourier transforming this result one finds (see appendix \ref{FT-greens-function} for details)
\begin{align}
\label{scalar-prop-mom}
\tilde{G}_R(p,p') = \tilde{G}_R^0(p) (2\pi)^4\delta^{(4)}(p-p') + \tilde{G}_R^0(p) \mathcal{T}(p,p')\tilde{G}_R^0(p')~,
\end{align}
where the effective vertex $\mathcal{T}$ is 
\be
\label{T-matrix}
\mathcal{T}(p,p') = - 4\pi i (p')^- \delta(p^- -(p')^-) \int d^2\bsz ~e^{i(\bsp-\bsp')\cdot \bsz}  \(e^{i f_1(\bsz)p'_+}-1\)~,
\ee
and $\tilde{G}^0_R(p)$ is given in Eq.~\eqref{free-G0-prop}. 
Just as in the gauge theory case, the form of the shock wave propagator in Eq.~\eqref{scalar-prop-mom} makes transparent the physical interpretation of multiple scatterings off the shock wave. The case of no interaction is captured by the first term in Eq.~\eqref{scalar-prop-mom} whereas the multiple interaction with the shock wave is captured by the $\mathcal{T}(p,p')$ function which resums all such interactions into an exponential (see the pictorial representation in Fig. \ref{gluon-shock wavepropagator}).

This result for the $\mathcal{T}(p,p')$ scalar amplitude was previously obtained by Gioia and Raclariu in \cite{deGioia:2022fcn} using a Bogoliubov transformation of the shock wave. As we see, this quantity is not the off-shell scalar propagator. Instead, as discussed for instance in \cite{Gelis:2006yv}, $\mathcal{T}(p,p')$ is the so-called retarded $T$ matrix (simply related to the Feynman amplitude multiplicatively by the matrix $U$), which gives a contribution to the NLO multiplicity\footnote{The other NLO mcontribution, as we will discuss later, is a one-loop correction to the scalar field in the shock wave background. For a detailed discussion of the relation between the Feynman and retarded amplitudes in the illustrative example of quark pair production in a gluon shock wave background, see for instance 
\cite{Blaizot:2004wv}.}:
\begin{equation}
\label{eq:on-shell-amplitude}
    \langle N\rangle_{\rm NLO (1)}= \int \frac{d^3 p}{(2\pi)^3 2\,E_p}\int\frac{d^3 q}{(2\pi)^3 2\,E_q} |\mathcal{T}(-q,p)|^2 \,.
\end{equation}


\subsection{Graviton shock wave propagators}
We move now to the computation of the graviton propagator in a  gravitational shock wave background. For a spin-2 propagator, the appropriate generalization of Eq.~\eqref{propDef1} and \eqref{gaugePropDef1} is (we drop the subscript $R$ in this sub-section to avoid clutter in the formulas):
\begin{align}
\label{gravPropDef1}
    G_{\m\n\rho\sigma}(x, y)=  -\int \frac{d^4k}{\(2\pi\)^4} \frac{1}{k^2+i\epsilon k^-}\sum_{\lambda\lambda'} h_{\m\n, k}^{(\lambda\lambda')}(x)h_{\rho\sigma, k}^{*(\lambda\lambda')}(y)~,
\end{align}
where the plane wave basis in the asymptotic past is given by
\be
h_{\m\n, k}^{(\lambda\lambda')}(x) = \epsilon^{(\lambda\lambda')}_{\m\n}(k) e^{-ikx} ~,\quad \text{(in the region } x^-<0)\,.
\ee
Here $\epsilon^{(\lambda\lambda')}_{\m\n}$ is the graviton polarization tensor which satisfies the completeness relation \cite{deAquino:2011ix}, 
\be
\label{polSum1}
\sum_{\lambda\lambda'} \epsilon^{(\lambda\lambda')}_{\m\n}(k) \epsilon^{*(\lambda\lambda')}_{\rho\sigma}(k) =\frac12 \(\Lambda_{\m\rho}\Lambda_{\n\sigma}+\Lambda_{\m\sigma}\Lambda_{\n\rho}-\Lambda_{\m\n}\Lambda_{\rho\sigma} \)~, \qquad \Lambda_{\m\n} = \eta_{\m\n} -\frac{n_\m k_\n + n_\n k_\m}{n\cdot k}~.
\ee
Here $n^\m$ is an arbitrary light-like vector. 
One  way to see the origin of this polarization sum  is to utilize the double-copy of the polarization vectors. Indeed, the physical graviton polarizations can be constructed\footnote{This is a convenient trick that extracts the physical graviton polarization from the ``Fat Graviton" double-copy field given by the naive tensor product of gluon polarizations. We thank J.J. Carrasco for a discussion on this point.} in terms of the gluon polarization tensor $\epsilon_\mu^{(\lambda)}$:
\be
\label{polarization-double-copy}
\epsilon_{\m\n}^{(\lambda\lambda')} = \frac{1}{2}\(\epsilon_\mu^{(\lambda)} \epsilon_\nu^{*(\lambda')}+\epsilon_\nu^{(\lambda)} \epsilon_\mu^{*(\lambda')} - \epsilon_\mu^{(\omega)} \epsilon_\nu^{*(\omega)}\delta^{\lambda\lambda'}\)~.
\ee
Using $k^\mu \epsilon_\m^{(\lambda)}=0$ and $\eta^{\m\n}\epsilon_\mu^{(\lambda)} \epsilon_\nu^{*(\lambda')}=
\delta^{(\lambda\lambda')}$, we see that the graviton polarization tensor constructed above satisfies the conditions of transversality and tracelessness: $k^\m\epsilon_{\m\n}^{(\lambda\lambda')}=\eta^{\m\n}\epsilon_{\m\n}^{(\lambda\lambda')} =0$. Furthermore, if the gluon polarization is in the light cone gauge ($n^\mu\epsilon_\m^{(\lambda)}=0$ for the null vector $n^\mu$), thus too is the graviton polarization. With this construction and the completeness relation for gluon polarization Eq.~\eqref{gauge-pol-completenessrel}, one can show that result in Eq.~\eqref{polSum1} follows straightforwardly.

Substituting the polarization sum in Eq.~\eqref{polSum1} into Eq.~\eqref{gravPropDef1}, we obtain
\begin{align}
\label{free-graviton-propagator}
    G^0_{\m\n\rho\sigma}(x, y) = -\frac12 \int \frac{d^4k}{\(2\pi\)^4} \frac{\(\Lambda_{\m\rho}\Lambda_{\n\sigma}+\Lambda_{\m\sigma}\Lambda_{\n\rho}-\Lambda_{\m\n}\Lambda_{\rho\sigma} \)}{k^2+ik^-\epsilon} e^{-ik(x-y)}~.
\end{align}

With this expression for the free propagator, we will now proceed with the computation of the graviton propagator in the background of the gravitational shock wave. Since we work in the eikonal approximation, the computation proceeds similarly to the scalar case. In  light cone gauge, it is sufficient to work with just the transverse components of the propagator since the remaining components are constrained by the relation $k^\mu G_{\m\n\rho\sigma}(k,k')=0$. The equations satisfied by the small fluctuation fields $h_{ij}(x)$ and the graviton propagator are given by\footnote{For completeness, the equations for the other components of the metric fluctuations are \cite{Raj:2023irr}:
\begin{align}
\label{graviton-eom-other}
& \square h_{i-}-g_{--} \partial_{+}^2 h_{i-}=\partial_{+} h_{i j} \partial_j g_{--}, \qquad \square h_{--}-g_{--} \partial_{+}^2 h_{--}=\left(\partial_i \partial_j g_{--}\right) h_{i j}+2\left(\partial_i g_{--}\right) \partial_j h_{i j} .
\end{align}
These equations mix the purely transverse components of the metric fluctuations $h_{ij}$ with the longitudinal ones: $h_{i-}$ and $h_{--}$. 
}
\begin{align}
\label{graviton-eom}
&2\p_+\p_- h_{i j}(x)-\square_\perp h_{i j}(x)-g_{--}(x) \partial_{+}^2 h_{i j}(x)=0~,\\[5pt]
\label{graviton-prop-eom}
&2\p_+\p_- G_{i j k l}(x,y)-\square_\perp G_{i j k l}(x,y)-g_{--} \partial_{+}^2 G_{i j k l}(x,y)=\frac12 \(\delta_{ik}\delta_{jl}+\delta_{il}\delta_{jk}-\delta_{ij}\delta_{kl} \) \delta^{(4)}(x-y)~.
\end{align}

Using the methods of appendix \ref{sec:AppB}, and the fact that $h_{ij}\delta_{ij}=0$ (c.f. section III B in \cite{Raj:2023irr}), it is straightforward to show that the retarded propagator $G_{ijkl}(x,y)$ satisfies the recursive relation
\begin{align}
\label{grav-prop-rec}
    G_{ijkl}(x,y) = \int d^4 z~ G_{ijmn}(x, z)\, \delta\left(z^{-} - z^{-}_0\right)\, 2 \,\partial_{z^+} G^{mn}_{~~~kl}(z,y)~,
\end{align}
where the contraction of the indices is performed using the Kronecker delta $\delta_{ij}$.

We will now use this property, as previously, to derive the result for the dressed graviton Green function. As previously noted, graviton polarizations are not affected due to the eikonal scattering off the shock wave. Applying Eq.~\eqref{grav-prop-rec} twice gives, 
\be
\label{three-term-relation-1}
G_{ijkl}(x,y) = \int d^4z \int d^4w ~G_{ijmn}(x, z) \delta\left(z^{-}-z_0^-\right) 2 \partial_{z^+} G^{mn}_{~~~pq}(z, w) \delta\left(w^{-}-w_0^-\right) 2 \partial_{w^+} G^{pq}_{~~~kl}(w,y)\,.
\ee
We choose again $x^->\delta$, $z_0^-=\delta$, $w_0^-=0$ and $y^-<0$, which allows us to rewrite the equation as  
\be
\label{three-term-relation-2}
G_{ijkl}(x,y) = \int d^4z \int d^4w ~G_{ijmn}^0(x, z) \delta\left(z^{-}-\delta\right)\, 2 \,\partial_{z^+} G^{mnpq}(z, w) \delta\left(w^{-}\right) \,2 \,\partial_{w^+} G^0_{pqkl}(w,y)\,,
\ee
where the free graviton propagators are given in Eq.~\eqref{free-graviton-propagator}. Note that when all the indices are in the transverse direction, these are proportional to the scalar propagator,
\be
G_{ijkl}^0(x, y) = \frac12\(\delta_{ik}\delta_{jl}+\delta_{il}\delta_{jk}- \delta_{ij}\delta_{kl} \) G^0_R(x,y)~,
\ee
with $G^0_R(x,y)$ defined in Eq.~\eqref{free-retarded-scalar-propagator}. 

As for the scalar case, the nontrivial term in Eq.~\eqref{three-term-relation-2} is $G^{mnpq}(z, w)$ which contains all the information about the interaction with the gravitational shock wave. To compute this term, we will need the the solution to the small fluctuation equations. We computed these previously in \cite{Raj:2023irr}, with the result
\begin{align}
    \label{grav-full-solution}
    h_{\m\n, k}^{(\lambda\lambda')}(x^-, x^+,\bsx) = e^{(\lambda\lambda')}_{\m\n}(k) \bigg[ \Theta(-x^-)\, e^{-ikx} + \Theta(x^-)\, e^{-ikx} U_k(\bsx) \bigg]~,
\end{align}
where $U_k(\bsx)$ was defined in Eq.~\eqref{grav-WL}.  
Inserting this expresion into Eq.~\eqref{gravPropDef1} gives
\be
\lim_{\delta \to 0}G_{ijkl}(z^-=\delta, w^-=0) = \frac14 \(\delta_{ik}\delta_{jl}+\delta_{il}\delta_{jk}- \delta_{ij}\delta_{kl} \) e^{-f_1(\bsz) \p_{z^+}}\Theta(z^+-w^+)~ \delta^{(2)}(\bsz-\bsw)\,. 
\ee
The novel feature relative to the scalar case is the overall tensorial factor $\frac{1}{2}(\delta_{ik}\delta_{jl}+\delta_{il}\delta_{jk}- \delta_{ij}\delta_{kl} ) $ which follows from the summation over graviton polarizations. Plugging this expression into Eq.~\eqref{three-term-relation-2} gives us the result for the shock wave propagator:
\begin{align}
    \label{gravitonPropPosFinal}
    G_{ijkl}(x,y) &= \frac12\(\delta_{ik}\delta_{jl}+\delta_{il}\delta_{jk}- \delta_{ij}\delta_{kl} \) \int d^4z ~G_R^0(x,z)\, \delta(z^{-}) \,e^{-f_1(\bsz)\p_{z^+}} \,2\,\p^-_z G_R^0(z,y)~.
\end{align}
The Fourier transform of this result can be expressed analogously to Eq.~\eqref{scalar-prop-mom} as
\begin{align}
\label{grav-prop-mom}
\tilde{G}_{ijkl}(p,p') = \tilde{G}_{ijkl}^0(p) (2\pi)^4\delta^{(4)}(p-p') + \tilde{G}_{ijpq}^0(p) \mathcal{T}^{pqrs}(p,p')\tilde{G}_{rskl}^0(p')~,
\end{align}
with $\mathcal{T}^{pqrs}(p,p')$ given by
\be
\label{T-matrix-1}
\mathcal{T}(p,p') = - \(\delta_{ik}\delta_{jl}+\delta_{il}\delta_{jk}- \delta_{ij}\delta_{kl} \)2\pi i (p')^- \delta(p^- -(p')^-) \int d^2\bsz ~e^{i(\bsp-\bsp')\cdot \bsz}  \(e^{i f_1(\bsz)p'_+}-1\)~.
\ee
Finally, we note that Eq.~\eqref{gravitonPropPosFinal} can be straightforwardly generalized to the fully covariant version\footnote{This is because  terms in the small fluctuation equations of motion that mix different components are suppressed at large transverse separations. This is evident from Eq.~\eqref{graviton-eom-other} where one observes that these terms are  proportional to $\p_i g_{--}\sim 1/b$ for large transverse separation $b$ between in the initial and final points $x,y$ in $G_{\m\n\rho\sigma}(x,y)$.},   
\begin{align}
\label{gravitonPropPosFinal2}
\tilde{G}_{\m\n\rho\sigma}(p,p') = \tilde{G}_{\m\n\rho\sigma}^0(p) (2\pi)^4\delta^{(4)}(p-p')+ \tilde{G}_{\m\n\alpha\beta}^0(p) \mathcal{T}^{\a\b\gamma\delta}(p,p')\tilde{G}_{\gamma\delta\rho\sigma}^0(p')~,
\end{align}
with $\mathcal{T}_{\m\n\rho\sigma}$ given by 
\begin{align}
\label{T-matrix-graviton}
    \mathcal{T}_{\m\n\rho\sigma}(p,p') = -\frac12 \(\Lambda_{\m\rho}\Lambda_{\n\sigma}+\Lambda_{\m\sigma}\Lambda_{\n\rho}-\Lambda_{\m\n}\Lambda_{\rho\sigma} \)4\pi i (p')^- \delta(p^- -(p')^-) \int d^2\bsz ~e^{i(\bsp-\bsp')\cdot \bsz}  \(e^{i f_1(\bsz)p'_+}-1\)~.
\end{align}
Eq.~\eqref{gravitonPropPosFinal2} is our final result for the graviton eikonal propagator in the shock wave background.  


\subsection{Shock wave propagator double-copy}
With these results, we can now compare the expressions for the 
retarded shock wave propagators in GR and in YM theory.  We first consider the colored scalar propagator in Eq.~\eqref{scalar-gauge-prop-mom} with Eq.~\eqref{scalar-prop-mom}. Apart from the structural similarity between these results, we can see that there is a classical double-copy~\cite{Monteiro_2014,Goldberger:2016iau,Goldberger:2017frp,White:2017mwc} relation between the two through their respective $\mathcal{T}$ amplitudes-inherited from the classical double-copy relation of the underlying shock wave backgrounds. The gravitational Wilson line appearing in Eq.~\eqref{grav-WL} can be written as 
\begin{align}
\label{grav-WL-1}
    U_k(\bsx) = \exp \bigg(i\kappa^2 \mu  \frac{\rho(\bsx)}{\square_{\perp}}k^- \bigg) = P\exp \left(\frac{i}{2} \int d z^{-} g_{--}\left(z^{-}, \bsx\right) k^-\right)~.
\end{align}
This expression should be compared to the Wilson line appearing for colored scalars in Eq.~\eqref{eq:QCD-Wilsonline}. The shock wave background has the double-copy $gA_- \to -\frac12 g_{--}$ previously noted in section III.C.4 of \cite{Raj:2023irr}). Under the replacement of the color charge density $\rho(\bsx)$ in YM to the mass density $\mu\,\rho(\bsx)$ in GR, $g\to\kappa$ and $T^a \to -k^-$, the QCD Wilson line operator maps exactly on to the gravitational Wilson line operator~\cite{Melville:2013qca}. This mapping is the classical double-copy relation between the retarded propagators of scalar fields in YM and in gravitational shock wave backgrounds. 

A similar comparison can be made between the results for retarded gluon and graviton shock wave propagators in Eq.~\eqref{gluonPropPosFinal2} and Eq.~\eqref{gravitonPropPosFinal2} respectively. The free retarded propagators in these expressions are satisfy the  to the differences in the tensor prefactors. As discussed previously, these too satisfy the double-copy
relation given in Eq.~\eqref{polarization-double-copy}. For the interacting terms, the double-copy relation between the respective $\mathcal{T}(p,p')$ amplitudes is identical to the relation discussed in the scalar case above. The reason is because the tensor prefactors are projectors 
\begin{equation}
\Lambda_{\mu\rho}\Lambda^{\rho}_{~\nu}=\Lambda_{\mu\nu}\qquad {\rm and}\qquad\Lambda_{\mu\nu\rho\sigma}\Lambda^{\rho\sigma}_{~~\alpha\beta}=\Lambda_{\mu\nu\alpha\beta}\qquad {\rm where} \qquad\Lambda_{\mu\nu\rho\sigma}=\frac12 \(\Lambda_{\m\rho}\Lambda_{\n\sigma}+\Lambda_{\m\sigma}\Lambda_{\n\rho}-\Lambda_{\m\n}\Lambda_{\rho\sigma}\) \,,
\end{equation}
which reduces the propagator for the graviton to that of the scalar propagator times the tensor factor. 


\subsection{Sub-eikonal corrections}
Our discussion thus far focused on the computation of eikonal propagators in shock wave backgrounds. In doing so, we neglected terms containing transverse derivatives acting on the mass distribution $\rho(\bsx)$. These terms are sub-eikonal as we will demonstrate shortly. For the case of scalar propagators (that lack physical polarizations), the generalization to include such sub-eikonal terms is straightforward. The steps leading to Eq.~\eqref{scalar-full-solution} will still be the same, except that now the gravitational Wilson line will contain extra terms with transverse derivatives acting on $\rho(\bsx)$:
\begin{align}
\label{grav-WL-full}
    U_k(x^-,\bsx) = \exp i\bigg( k^- f_1(\bsx)+ k^- x^-f_2(\bsx) -x^- k_i f_{3,i}(\bsx) \bigg)~,
\end{align}
where the functions $f_1, f_2, f_{3,i}$ are given by
\begin{align}
    f_1(\bsx) =  \kappa^2 \mu  \frac{1}{\square_{\perp}}\rho(\boldsymbol{x})~, \qquad f_2(\bsx) = -\frac{1}{2} \kappa^4 \mu^2 x^{-} \left(\frac{\partial_i}{\square_{\perp}} \rho(\boldsymbol{x})\right)^2~, \qquad f_{3,i}(\bsx) = \kappa^2 \mu x^{-}  \frac{\partial_i}{\square_{\perp}} \rho(\boldsymbol{x})~.
\end{align}
The term $f_1$ was the eikonal piece we computed previously and the other two expressions are the sub-eikonal contributions. 

The result in Eq.~\eqref{grav-WL-full} for the Wilson line follows from the solution to the geodesic equations given in Eq.~\eqref{xytransformation}. A scalar wave packet prepared in the region $x^-<0$ will traverse along a geodesic path prescribed by the above formula \cite{Wald:1984rg}. In other words, the full solution to the small fluctuation scalar equations can be written as in Eq.~\eqref{scalar-full-solution} where the Wilson line $U_k(x^-,\bsx)$ is given by Eq.~\eqref{grav-WL-full}. This result for the small scalar fluctuations can then be used to compute the retarded scalar propagator (following Eq.~\eqref{non-trivial-prop} onwards) having all the sub-eikonal terms, captured by $f_3$ and $f_2$ taken into account. The latter is seen from the fact $f_3$ and $f_2$ are suppressed w.r.t. $f_1$ by powers of $R/b$ and $(R/b)^2$ respectively where the characteristic length scale $R$ is given by $R\sim \kappa^2 \mu$ and we used the fact that for large impact parameters we can approximate $\rho(\bsx)/\square_\perp \sim \log(b)$. (Recall that $\mu$ is the energy of the shock wave appearing as a source term in the Einstein's equations of motion $T_{\m\n}=\mu \delta_{-\m}\delta_{-\n}\delta(x^-)\rho(\bsx)$.) 

In the previous section, we ignored the effect on the graviton polarization coming from the propagation through the shock wave. These effects are sub-eikonal and are more complicated to work out. As previously noted, for $x^-<0$, the spin 2 wave packet is given by
\be
\label{spin2basisbefore}
h_{\m\n, k}^{(\lambda\lambda')}(x^-<0,x^+,\bsx) = e^{(\lambda\lambda')}_{\m\n}(k)\, e^{-ikx}\, \Theta(-x^-)~.
\ee
In order to compute the propagator, we need to determine how these basis elements change in $x^->0$ region. The analysis of the $e^{-ikx}$ terms was performed above. To understand how the polarization vectors evolve as the wave packet traverses the gravitational shock wave is to simply transform the spin-2 fluctuations along null geodesics crossing the shock wave, given by the coordinate transformation of a rank-two tensor,
\begin{equation}
h_{\mu \nu}(y)=\frac{\partial x^\rho}{\partial y^\mu} \frac{\partial x^\sigma}{\partial y^\nu} h_{\rho \sigma}(x)~,
\end{equation}
where the transformation relations $x\equiv x(y)$ were given in Eq.~\eqref{xytransformation}. For instance, if we look at how the transverse components of the graviton behave, we find
\be
h_{ij}(y)=\frac{\partial x^+}{\partial y^i} \frac{\partial x^+}{\partial y^j} h_{++}(x) + \(\frac{\partial x^+}{\partial y^i} \frac{\partial x^k}{\partial y^j} +\frac{\partial x^k}{\partial y^i} \frac{\partial x^+}{\partial y^j} \)h_{+k}(x) + \frac{\partial x^k}{\partial y^i} \frac{\partial x^l}{\partial y^j} h_{kl}(x)\,.
\ee
Inspecting Eq.~\eqref{xytransformation}, we find that upon ignoring all terms which have a transverse derivative acting on $\rho(\bsx)$, one gets  
\be
h_{ij}(y) \sim \frac{\partial x^k}{\partial y^i} \frac{\partial x^l}{\partial y^j} h_{kl}(x) \sim \delta_i^k \delta_j^l h_{kl}(x) = h_{ij}(x)~.
\ee
This result is in line with the expectation that the polarization tensor is not affected at leading eikonal order. Similar results hold for the other components of the graviton field. These results confirm that the graviton polarization tensor, at leading eikonal order, is not modified by scattering off the shock wave. The calculation of the propagator is therefore essentially the same as that of the scalar propagator. We leave a systematic treatment of the sub-eikonal corrections sketched above to future work.

\section{Outlook: A shock wave road map to the gravitational BFKL equation}

In this section, we will discuss our primary motivation for computing the gravitational shock wave propagators and sketch out their role in gravitational radiation at high frequencies. Before we do so, we will first briefly provide the ``lay of the land" in this context. The $2\to2$ S-matrix in gravitational scattering is frequently expressed as~\cite{Amati:1990xe} 
\begin{equation}
    S= e^{2i(\delta_0+\delta_1+\delta_2+\cdots)}\,,
\end{equation}
where the leading eikonal term including multiple {\it soft} graviton exchange is 
\begin{equation}
    \delta_0=G s\ln\(\frac{L}{b}\)\,,
\end{equation}
where $L$ is an infrared cutoff and $b$ is the impact parameter. 
In this power counting,
\begin{equation}
    \delta_1 = \frac{6 G^2 s}{\pi b^2}\, \ln(s)\,,
\end{equation}
is a quantum gravity correction, which is power suppressed by $\sim Gs(L_p^2/b^2)$, where $L_p$ is the Planck length.  Both of these contributions are purely real and contribute only to elastic scattering. The first inelastic contribution comes in at the next (2-loop) order in $2\rightarrow 2$ scattering:
\begin{equation}
    \delta_2= \frac{2 G^3 s^2}{b^2}\left[1+\frac{i}{\pi}\ln(s)\left(\ln\frac{L^2}{b^2}+2\right)\right]\,.
\end{equation}
In Regge asymptotics, this is the so-called H-diagram of Amati, Ciafaloni and Veneziano~\cite{Amati:1990xe} , where the Lipatov vertex first appears. 

It is the absorptive part of this 2-loop contribution which provides the leading order contribution to gravitational radiation in the ``deeply inelastic" 
scattering regime that we computed in \cite{Raj:2023irr}.  It is the leading contribution when $b$ approaches $R_S$ but is still sufficiently large that many-body inelastic graviton interactions are suppressed. The shock wave propagators are crucial in computing the leading NLO contributions (three-loops in the usual power counting) to the H-diagram, and in their resummation to all-orders. 
This resummation was first performed by Lipatov~\cite{Lipatov:1982it,Lipatov:1982vv} using Feynman diagram techniques, and he obtained the leading result for the $2\rightarrow N$ cross-section to be 
\begin{equation}
    \sigma(s)= \frac{{\rm Im} \,M_{2\rightarrow 2}}{s}= s^2 \int \frac{d\omega}{2i}\,s^\omega f_\omega(k)|_{k^2=0,q^2\rightarrow 0} \,.
\end{equation}
Here ${\rm Im} \,M_{2\rightarrow 2}$ is the imaginary part of the $2\rightarrow 2$ forward scattering amplitude, and the amplitude $f_\omega$ in Mellin space satisfies a Bethe-Salpeter-type integral equation 
\begin{equation}
\label{eq:Gravi-BFKL}
    \left[\omega -2\,\omega(k^2)\right]f_\omega(k) = \frac{\kappa^2}{q^4}+\kappa^2\int \frac{d^2 q_\perp}{2(2\pi)^3}\,\frac{1}{q_\perp^4}\, K(k,q)|_{q^2\rightarrow 0}\, f_\omega(q)\,.
\end{equation}
In this expression, $\omega(k^2)$ is the so-called graviton Reggeon trajectory, which satisfies the equation~\cite{Lipatov:1982it,Lipatov:1982vv,Lipatov:1991nf} 
\begin{equation}
\label{eq:Regge-trajectory}
    \omega(k^2) = k^2  \frac{\kappa^2}{(2\pi)^3}\,\int\frac{d^2 q_\perp}{q_\perp^2 (k_\perp-q_\perp)^2}\,\left[({\bf q}_\perp\cdot({\bf k}_\perp - {\bf q}_\perp))^2\left(\frac{1}{q_\perp^2}+\frac{1}{(k_\perp-q_\perp)^2}\right) -  k^2 \right]\approx  
    -\frac{\kappa^2}{(2\pi)^3}\,k^2\ln\left(\frac{k_\perp^2}{\Lambda^2}\right)\,,
\end{equation}
with the last expression being valid when $k_\perp \ll q_\perp$, and 
where $\Lambda$ is an infrared cutoff scale. Further, $K(k,q)= \Gamma^{\mu\nu}(k,q) \Gamma_{\mu\nu}(-k,-q)$ is simply the scalar product of the gravitational Lipatov vertex which, as we noted in Eq.~\eqref{grav-lipatov-vertex}, is itself a bilinear of a combination of the QCD Lipatov vertex and the photon bremsstrahlung vertex. 

The construction sketched here for the computation of the $2\rightarrow N$ amplitude is precisely that followed by Lipatov and collaborators\footnote{See \cite{DelDuca:1995hf,DelDuca:2022skz} for excellent reviews of the BFKL pomeron.} in their computation of the BFKL pomeron in QCD, understood to be the color singlet compound state of two reggeized gluons~\cite{Kuraev:1977fs,Balitsky:1978ic}.  One can similarly think of Eq.~\eqref{eq:Gravi-BFKL} as the gravitational BFKL equation, with the ``pomeron" in this case, understood as a compound state of two reggeized gravitons~\cite{Lipatov:1982vv}. An important subtlety that modifies the above discussion is in the treatment of ultraviolet divergences at one-loop. As pointed out in 
\cite{Bartels:2012ra}, the one-loop correction to the $2\rightarrow 2$ Born amplitude (responsible for the double logs $\ln(k^2/\Lambda)\,\ln(s/k^2)$ contributing to the Regge trajectory $\omega(k^2)$) can be expressed as 
\begin{equation}
    M^{(1) {\rm Regge}}_{2\rightarrow 2}\sim \frac{1}{\pi}\int \frac{d^2 q_\perp}{q_\perp^2}\,\int_{k^2}^s \frac{d(\beta s)}{\beta s-q_\perp^2}\,,
\end{equation}
where $0\leq \beta \leq 1$. This expression contributes an additional ``Sudakov" double logarithmic contribution $\ln^2(s/k^2)$ to the original Lipatov result. One can understand this as a finite energy constraint on the evolution - which must be strictly imposed when the relevant current is the stress-energy tensor. Interestingly, similar double logarithmic constraints appear in QCD as well when one imposes finite energy constraints on the bremsstrahlung logs in $x$ and $k_\perp$~\cite{Salam:1998tj}, leading to a kinematically constrained BFKL evolution\footnote{For a recent discussion in the context of back-to-back QCD jets in DIS at NLO, see \cite{Caucal:2023fsf,Caucal:2023nci}.}.

\begin{figure}[ht]
\centering
\includegraphics[scale=1.0]{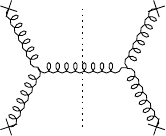}
\caption{Cut vacuum-to-vacuum H diagram contributing at leading order to Eq.~\eqref{leading-order}. The crosses depict the source densities, which are often depicted as ``sewn" together. }
\label{h-diagram}
\end{figure}

\begin{figure}[ht]
\centering
\raisebox{-32pt}{\includegraphics[scale=1.0]{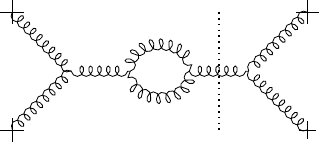}}
+
\raisebox{-32pt}{\includegraphics[scale=1.0]{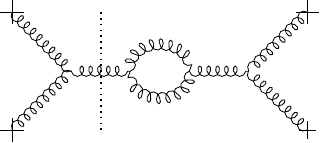}}
+
\raisebox{-32pt}{\includegraphics[scale=1.0]{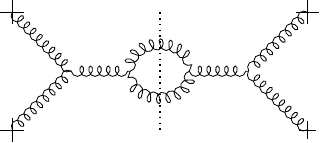}}
\caption{These cut vacuum-vacuum diagrams depict both higher order insertions of the classical field satisfying $(g\rho)^n$ expansion as well next-to-leading order $O(g^2)$ corrections to the $O(1)$ leading order contribution shown in Fig.~\ref{h-diagram}. Large rapidity logarithms ($g^2 Y)$ requiring further resummation are generated at NLO.}
\label{NLO-contributions}
\end{figure}

As outlined in the introduction, one of the goals of our work is to obtain in the shock wave scattering picture the aforementioned state-of-the art in the computation of gravitational radiation in trans-Planckian scattering at high frequencies.  To go beyond the leading order result in \cite{Raj:2023irr}, a key step is the computation of the shock wave propagators performed in this work. Indeed, one of the NLO contributions to the inclusive spectrum is the expression in 
Eq.~\eqref{eq:on-shell-amplitude}. As discussed in \cite{Gelis:2008rw}, this contribution corresponds to the contribution from the cut ``Wightman" propagator $G_{+-}$. The other NLO contribution is from the interference term between the leading order contribution and the one-loop correction to the leading order expression. This latter contribution can be computed from the Feynman propagator $G_{++}$. The leading order cut H-diagram is shown in Fig.~\ref{h-diagram} and both NLO contributions  are shown in Fig.~\ref{NLO-contributions}.

However as shown previously for gluon radiation in shock wave scattering in the CGC EFT, the NLO contributions to the radiation spectrum can be formulated entirely as an initial value problem with the key ingredient in the computation being the retarded dressed Green function in the shock wave background. For illustration, we  will sketch this discussion in the QCD case below. As we have seen, since the gravitational propagators have an analogous structure, we expect the same logic to hold.

 We will consider the observable \cite{Gelis:2007kn,Gelis:2008rw}
\be
\label{inclusive-observables}
\mathcal{O}^{ij}(x,y) \equiv \<A^i(x)A^j(y)\>~,
\ee
the Fourier transform of whose trace corresponds to the single inclusive number distribution of gluons produced in a shock wave collision. In general, this computation of products of out-of-time ordered fields can be formulated in terms of ``vacuum-to-vacuum" diagrams in the presence of strong external sources. In the SK ``in-in" formalism, the field $A^i(x)$ is placed on the $(-)$ branch and $A^j(y)$ on the $(+)$ branch of the Schwinger-Keldysh contour; this construction is discussed further in appendix~\ref{sec:AppA}.

At leading  leading order in the coupling expansion, Eq.~\eqref{inclusive-observables} is simply the product of the classical solution to the YM equations in the shock wave background, 
\be
\label{leading-order}
\mathcal{O}^{ij}_{\mathrm{LO}}(x, y)=\mathcal{A}^i(x) \mathcal{A}^j(y)~,
\ee
where $\mathcal{A}$ is the solution to the classical equations of motion that vanishes in the far past
\begin{align}
{\left[\mathcal{D}_\nu, \mathcal{F}^{\mu \nu}\right] } =J^\nu~,\qquad\qquad \lim _{x^0 \rightarrow-\infty} \mathcal{A}^\mu(x) =0~.
\end{align}
Diagrammatically, the way one sees this is that for a given on-shell 
positive frequency $+$ field produced, one is summing over trees of graphs, where at each vertex one sums over $\pm$ vertices. Thus each Feynman propagator $G_{++}$ is replaced by $G_R=G_{++}-G_{+-}$, with $G_{+-}$ being the on-shell Wightman function. 

The same thing happens at NLO in the power counting~\cite{Gelis:2006cr,Gelis:2006ye}. The NLO corrections to  inclusive gluon radiation in shock wave collisions can be expressed as~\cite{Gelis:2008rw} 
\begin{equation}
\label{NLO-correction}
\mathcal{O}^{ij}_{\text {NLO}}(x, y)=\mathcal{A}^i(x) \beta^j(y)+\mathcal{A}^j(y)\beta^i(x) +\mathcal{G}^{ij}_{-+}(x, y)~.
\end{equation}
Here $\mathcal{A}^i(x)$ denotes the classical field created by the gluon shock wave and $\beta^i(x)$ is the one-loop correction to the classical field. These diagrams are analogous to those shown in Fig.~\ref{NLO-contributions}. 

In the SK approach,  $\beta^i(x)$ satisfies a wave equation in the presence of the classical background with the source term 
\be
\mathcal{G}_{++}^{i j, b c}(x, x)=\sum_{\lambda, a} \int \frac{d^3 \boldsymbol{k}}{(2 \pi)^3 2 E_{\boldsymbol{k}}} a_{-\boldsymbol{k} \lambda a}^{i b}(x) a_{+\boldsymbol{k} \lambda a}^{j c}(x)~.
\ee
Here $\lambda$ is the gluon polarization and $a,b,c$ are adjoint color indices. Specifically, the fields $a_{+\boldsymbol{k} \lambda a}^{j c}(x)$ are plane waves at initial times with initial color $a$ and final color $c$;  the subscript $\pm$ indicates the fields on the two SK contours. To obtain $\beta$, one needs to solve the aforementioned small fluctuation equation in the presence of the background field $\mathcal{A}$ and the source term $G_{++}$ with retarded boundary conditions at $x^0\to - \infty$. The expression for the cut propagator $G_{+-}$ is analogous, except one solves the small fluctuation equations for $a_{+\boldsymbol{k} \lambda a}^{j c}(x)$ without the source term. 

As shown in \cite{Gelis:2008rw}, both NLO contributions in Eq.~\eqref{NLO-contributions} can be computed entirely in terms of retarded Green functions. This provides the most general treatment especially in the full-blown strong field regime of scattering. However away from it, one can work with either retarded or Feynman/Wightman dressed propagators (which as noted previously are simply related by a Wilson line) and compute the NLO $G_{++}$ and $G_{+-}$ contributions directly. These  gluon shock wave contributions to $\mathcal{O}$ can be recast into a form of an operator $\mathbb{H}$ acting on the leading order observable $\mathcal{O}_{\rm LO}$  (we suppress the Lorentz indices $i,j$ below) as 
\be
\label{QCD-phase-space-integral}
\mathcal{O}_{\rm NLO} = \mathbb{H}~[\mathcal{O}_{\rm LO}] + \Delta \mathcal{O}_{\rm NLO}\,,
\ee
where $\mathbb{H}$ is the BFKL Hamiltonian, which is proportional to 
$\alpha_S \ln(1/x)\sim 1$ for $x\ll 1$. The second term $\Delta \mathcal{O}_{\rm NLO}$ is free from such large logs and doesn't contribute to ``leading logarithmic accuracy". The solution of this BFKL RG equation resums  the leading $(\alpha_S\ln(1/x))^N$ contributions to the inclusive distribution $\mathcal{O}$ to all orders in perturbation theory. 

These results in QCD therefore indicate the crucial importance of the shock wave propagators both in recovering Lipatov's gravitational BFKL equation and in attempts to go further in the strong field regime where many-body rescattering effects become important\footnote{We note that such rescattering contributions have been considered in the context of the leading order H-diagram previously~\cite{Ciafaloni:2015xsr}.}. Because of the analogous structure of eikonal shock wave propagators in QCD and GR, we expect the former correspondence to be straightforward. The latter is significantly more involved than the QCD case because it will require careful treatment of sub-eikonal contributions. These studies will be pursued separately. 

As a final note, we will comment on the implications of these studies for high frequency gravitational wave radiation. The spectrum of this radiation obtained within the related ACV framework can be found in \cite{Ciafaloni:2015vsa,Ciafaloni:2015xsr}. The gravitational spectrum $dE/d\omega$ in the low frequency region $\omega R_S\ll 1$ is flat. This changes qualitatively in the Lipatov regime $1<\omega R_S < \omega_M R_S$, where the high frequency cutoff $\omega_M\sim M_{\rm BH}/\hbar$ for black hole mass $M_{\rm BH}$. In this regime, 
$dE/d\omega\sim1/\omega$. An interesting question is what happens at impact parameters $b\leq R_S$. It has been argued that significant high frequency {\it coherent} gravitational radiation can occur in this regime~\cite{Ciafaloni:2017ort,Colferai:2022orh}. As a first step, we will confirm if the ACV results are recovered in our shockwave RG framework. As noted, we anticipate significant deviations from the ACV results when one enters the  strong field regime of high occupancies. We also plan to explore whether these modifications of the high frequency radiation spectrum can be measured at at future gravitational wave observatories. In particular it will be important to examine whether contributions from ultrarelativistic black hole scattering in the high occupancy regime will have characteristic signatures (see for example \cite{Dvali:2021ofp,Escriva:2022duf}), that are distinguishable from the stochastic gravitational wave background.

\begin{acknowledgments}
We are grateful to P. V. Athira, Leonardo de la Cruz, Lance Dixon, Nava Gaddam, Siddharth Prabhu and Ana-Maria Raclariu for helpful discussions. We would like to especially thank John Joseph Carrasco for valuable insights into double-copy relations. R.V is supported by the U.S. Department of Energy, Office of Science under contract DE-SC0012704 and within the framework of the SURGE Topical Theory Collaboration. He acknowledges partial support from an LDRD at BNL. R.V was also supported at Stony Brook by the Simons Foundation as a co-PI under Award number 994318 (Simons Collaboration on Confinement and QCD Strings). H.R is a Simons Foundation postdoctoral fellow at Stony Brook supported under Award number 994318. 
\end{acknowledgments}


\appendix

\section{Schwinger-Keldysh approach and connection to scattering amplitudes}
\label{sec:AppA}
The Schwinger-Keldysh formalism (or the in-in formalism) is a powerful framework for studying QFTs at finite temperature or coupled to external sources (for instance external shock wave background). It provides a systematic way to compute thermal averages, correlation functions and real-time dynamics of quantum fields in the presence of external sources. For an excellent modern treatment of the SK formalism in quantum field theory, we refer the reader to \cite{Berges:2004yj}. The key component of the Schwinger-Keldysh approach is the use of a closed time path contour comprised of two branches as shown in Fig.~\ref{SK_contour} below. The upper ($+$) branch runs along the real time axis in the positive direction, while the lower ($-$) branch runs backwards along the negative real axis. Operators and correlation functions are defined by path ordering along this closed time contour. This contour structure allows the formalism to naturally capture the impacts of external sources on the dynamics of quantum fields.
\begin{figure}[ht]
\centering
\includegraphics[scale=1.0]{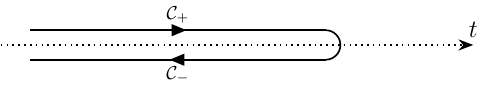}
\caption{The Schwinger-Keldysh contour closed-time contour.}
\label{SK_contour}
\end{figure}

To perform calculations, the Schwinger-Keldysh formalism introduces special Feynman rules adapted to the closed time path. Integrals are performed over the full contour, and propagators are expressed as path-ordered products of asymptotic fields living on the two contour branches. The propagators split into four types ($++$, $--$, $+-$, $-+$) depending on which branches the fields reside on. These are defined as 
\begin{align}
\begin{split}
& G_{++}^0(x, y) \equiv\left\langle 0_{\mathrm{in}}\left|T \phi_{\text {in }}^{(+)}(x) \phi_{\text {in }}^{(+)}(y)\right| 0_{\mathrm{in}}\right\rangle~,\qquad G_{-+}^0(x, y) \equiv\left\langle 0_{\text {in }}\left|\phi_{\text {in }}^{(-)}(x) \phi_{\text {in }}^{(+)}(y)\right| 0_{\text {in }}\right\rangle \\
& G_{--}^0(x, y) \equiv\left\langle 0_{\mathrm{in}}\left|\bar{T} \phi_{\text {in }}^{(-)}(x) \phi_{\text {in }}^{(-)}(y)\right| 0_{\text {in }}\right\rangle ~,\qquad G_{+-}^0(x, y) \equiv\left\langle 0_{\text {in }}\left|\phi_{\text {in }}^{(+)}(y) \phi_{\text {in }}^{(-)}(x)\right| 0_{\text {in }}\right\rangle~.
\end{split}
\end{align}
Note that as opposed to the commonly used `in-out' formalism, here one uses only the `in' vacuum $| 0_{\text {in}}\rangle$. The free propagators as usual denoted with a superscript $0$. These quantities are inter-related as follows:
\begin{align}
\begin{split}
    &G_{++}^0(x, y) \equiv \theta\left(x^0-y^0\right) G_{-+}^0(x, y)+\theta\left(y^0-x^0\right) G_{+-}^0(x, y)~,\\
    &G_{--}^0(x, y) \equiv \theta\left(x^0-y^0\right) G_{+-}^0(x, y)+\theta\left(y^0-x^0\right) G_{-+}^0(x, y) ~,
\end{split}
\end{align}
which implies $G_{++}^0+G_{--}^0=G_{-+}^0+G_{+-}^0$. The momentum space expression of these quantities are
\begin{align}
& G_{++}^0(p)=\frac{i}{p^2+i \epsilon}, \quad G_{--}^0(p)=\frac{-i}{p^2-i \epsilon}~,\quad G_{-+}^0(p)=2 \pi \theta\left(p^0\right) \delta\left(p^2\right), \quad G_{+-}^0(p)=2 \pi \theta\left(-p^0\right) \delta\left(p^2\right)
\end{align}
The relation to retarded propagator is as follows
\begin{align}
G_R^0 = G_{++}^0-G_{+-}^0 = G_{-+}^0-G_{--}^0~.
\end{align}
In the presence of a classical background field the full Schwinger-Keldysh propagators $G_{+-}, G_{-+}$ (that resums all the background insertions) can be obtained from solving the Lippmann-Schwinger equations
\begin{align}
\label{full-sk-propagators}
& G_{-+}=G_R G_R^{0-1} G_{-+}^0 G_A^{0-1} G_A~,\qquad G_{+-}=G_R G_R^{0-1} G_{+-}^0 G_A^{0-1} G_A~.
\end{align}
where as we have already seen the full retarded (and also the advanced) propagators can be decomposed in terms of the retarded and advanced $\mathcal{T}$ matrices as: $ G_R \equiv G_R^0+G_R^0 \mathcal{T}_R G_R^0~, G_A \equiv G_A^0+G_A^0 \mathcal{T}_A G_A^0$. The Schwinger-Keldysh propagators $G_{-+}, G_{+-}$ are of fundamental importance in performing next-to-leading order computations. The key point is that the formulas Eq.~\eqref{full-sk-propagators} makes it possible to perform next-to-leading order computations (to all orders in the background field) by essentially solving equations of motion for the classical and small fluctuation fields with entirely retarded boundary conditions. See \cite{Gelis:2006yv} for a computation of the next to leading order average multiplicity $\<n\>$ in $\phi^3$ theory in the presence of strong background fields and \cite{Gelis_2008_I} for a derivation of the JIMWLK Hamiltonian in QCD that makes use of purely retarded Green’s functions.


\section{Convolution property of retarded Green's function}
\label{sec:AppB}
Here we derive a convolution property of the spin-0 shock wave Green's function. 
Consider the equation for the scalar field in the presence of a gravitational shock wave and an arbitrary source $J(x)$ and the corresponding equation for the retarded Green's function:
\begin{align}
    &2\p_+\p_-\phi(x)-g_{--}\p_+^2\phi(x)-\square_\perp\phi(x) = J(x)\,, \no\\[5pt]
    &2\p_+\p_-G_R(x, y)-g_{--}\p_+^2 G_R(x, y)-\square_\perp G_R(x, y) = \delta^{(4)}(x-y)\,,
\end{align}
where we note that $g_{--}$ is not a function of $x^+$. We multiply the first equation by $G(x,y)$ and the second equation by $\phi(x)$. We then integrate these equations and subtract them, which gives the  result for the field $\phi(x)$ to be 
\begin{align}
\label{eq-temp}
    \phi(x) = \int_{y^->y_0^-}d^4y~ G_R(x,y)J(y) + \int_{y^->y_0^-}d^4y~ \bigg(\phi(y) \mathcal{O}(y) G_R(x,y)-G_R(x,y)\mathcal{O}(y) \phi(y) \bigg)\,,
\end{align}
where the operator $\mathcal{O}(y)$ is
\be
\mathcal{O}(y) = 2\frac{\p}{\p y^+}\frac{\p}{\p y^-} -\nabla_y^2 - g_{--}(y)\frac{\p^2}{\p (y^+)^2}\,.
\ee
The integration regions in these formulae are taken to be $y^->y_0^-$, since we wish to determine the field in the region $y^->y_0^-$ with the initial condition given at $y^-=y^-_0$. We simplify the second integral as follows: 

\begin{itemize}
    
\item Term involving $\frac{\p}{\p y^+}\frac{\p}{\p y^-}$ :

\begin{align}
    &2\int_{y^->y_0^-}d^4y~ \bigg(\phi(y) \frac{\p}{\p y^+}\frac{\p}{\p y^-} G_R(x,y)-G_R(x,y)\frac{\p}{\p y^+}\frac{\p}{\p y^-} \phi(y) \bigg)\no\\[5pt]
    =&2\int_{y^->y_0^-}d^4y~ \bigg(\frac{\p}{\p y^+} \[\phi(y) \frac{\p}{\p y^-} G_R(x,y)\] - \frac{\p}{\p y^-} \[G_R(x,y)\frac{\p}{\p y^+} \phi(y)\] \bigg)\no\\[5pt]
    =&2\int_{y^->y_0^-}d^2\bsy dy^-  \[\phi(y) \frac{\p}{\p y^-} G_R(x,y)\]_{y^+=-\infty}^{y^+=\infty}  -2\int d^2\bsy dy^+  \[G_R(x,y)\frac{\p}{\p y^+} \phi(y)\]_{y^-=y_0^-}^{y^-=\infty}  \no\\[5pt]
    =&2\int d^2\bsy dy^+  \[G_R(x,y)\frac{\p}{\p y^+} \phi(y)\]_{y^-=y_0^-} = \int d^4y ~ G_R(x,y)\delta(y^--y_0^-)2\p_+ \phi(y)~.
\end{align}
In the last equation, we used the fact that the field $\phi(x)$ and its derivatives decays for asymptotic values of $y^+$ and $G_R(x,y)$ being a retarded propagator goes to zero as $y^-\to\infty$. 

\item Term involving $\nabla_y^2$ : For this term one can show using Stokes' theorem \cite{Itzykson:1980rh, Blaizot:2004wu} that the result depends on scalar field, its derivatives at asymptotic infinity in the transverse directions. Therefore this contribution vanishes identically. 

\item Finally for the term involving $g_{--}(y)\frac{\p^2}{\p (y^+)^2}$, manipulating the integral a bit we get
\begin{align}
    &\int_{y^->y_0^-}d^4y~ \frac{\p}{\p y^+}\bigg(\phi(y) g_{--}(y) \frac{\p}{\p y^+} G_R(x,y)-G_R(x,y)g_{--}(y)\frac{\p}{\p y^+} \phi(y) \bigg)\no\\[5pt]
    =&\int_{y^->y_0^-}d^2\bsy dy^- ~ \[\phi(y) g_{--}(y) \frac{\p}{\p y^+} G_R(x,y)-G_R(x,y)g_{--}(y)\frac{\p}{\p y^+} \phi(y) \]_{y^+=\infty}^{y^+=-\infty}
    \label{step-manipulation}
\end{align}
Again, since this term depends upon $\phi(x)$ and its derivatives at asymptotic values of $y^+$ we can drop this contribution altogether.
\end{itemize}
Hence, we get the final solution for the scalar field,
\begin{align}
    \phi(x) = \int_{y^->y_0^-}d^4y~ G_R(x-y)J(y) + \int d^4y ~ G_R(x,y)\delta(y^--y^-_0)2\p_+ \phi(y)~.
\end{align}
For vanishing source $J(x)$, the result is simply,
\begin{align}
    \phi(x) = \int d^4y ~ G_R(x,y)\delta(y^--y^-_0)2\p_+ \phi(y)~.
\end{align} 
The final step is the realization that the above relation is recursive. If we apply it twice we get 
\begin{align}
    \phi(x) = \int d^4y ~ G_R(x,y)\delta(y^--y^-_0)2\p_+ \int d^4z ~ G_R(y,z)\delta(z^--z^-_0)2\p_+ \phi(z)~.
\end{align}
Comparing this equation with the previous one (where we change the dummy variable $y$ to $z$ on the left side) we find 
\be
G_R(x,z) = \int d^4y ~ G_R(x,y)\delta(y^--y^-_0)2\frac{\p}{\p y^+} G_R(y,z)\,.
\ee
This establishes the relation in Eq.~\eqref{retarded-prop-conv} used in the main text for the retarded Green's function of scalar field. Note that this relation is generic and independent of the nature of shock wave background as is seen from Eq.~\eqref{step-manipulation}. It holds for both gluon and gravitational shock wave backgrounds. 

\subsection{Explicit expression for the retarded Green's function in position space}
\label{app:B2}
Here we give the details regarding the retarded scalar Green's function in position space written in Eq.~\eqref{free-retarded-scalar-propagator}. The momentum space representation of the Green's function is
\be
G_R(x) = -\int \frac{d^4k}{(2\pi)^4} \frac{e^{-ikx}}{k^2+ik^-\eps}\,.
\ee
Performing the integral using Cauchy's theorem,
\begin{align}
\begin{split}
    G_R(x) &= -\int \frac{dk^+}{2\pi}\frac{dk^-}{2\pi} \frac{d^2\bsk}{(2\pi)^2} ~\frac{e^{-ikx}}{2k^+k^--\bsk^2+ik^-\eps}\\[5pt]
    &= -\int \frac{dk^+}{2\pi}\frac{dk^-}{2\pi} \frac{d^2\bsk}{(2\pi)^2} ~\frac{e^{-ikx}}{2k^-\(k^+-\frac{\bsk^2}{2k^-}+i\frac{\eps}{2}\)}\\[5pt]
    &= i\Theta(x^-)\int \frac{dk^-}{2\pi 2k^-} \frac{d^2\bsk}{(2\pi)^2} ~\exp-i\(\[\frac{\bsk^2}{2k^-}-i\frac{\eps}{2}\]x^- +k^- x^+ - \bsx\cdot \bsk \)\\[5pt]
    &= i\Theta(x^-)\int \frac{dk^-}{2\pi 2k^-} \int_0^\infty \frac{du}{2\pi} ~u\exp\[-\frac{ix^-}{2k^-}u^2\]\exp\[-\frac{ik^-}{2x^-}x^2\]\\[5pt]
    &= -i\frac{1}{4\pi}\Theta(x^-)\int \frac{dk^-}{2\pi }  ~\frac{i}{x^-}e^{-i\frac{k^-}{2x^-} x^2}\\[5pt]
    &=\frac{1}{2\pi}\Theta(x^-)\Theta(x^+)\delta\(x^2\)
\end{split}
\end{align}

\subsection{Fourier transform of shock wave propagators}
\label{FT-greens-function}
Here we spell out the steps for arriving at the momentum space formula for the scalar shock wave propagator. Starting from the position space result in Eq.~\eqref{scalar-gauge-prop} and the definition Eq.~\eqref{fourier-transform-def}, we have
\begin{align}
    \tilde G_R(p,p') &= \tilde G_R^0(p,p') + \int d^4x ~d^4y~ e^{ipx}e^{-ip'y}\int d^4z ~G_R^0(x,z) \delta(z^{-}) \(U(\bsz)-1\) 2\p^-_z G_R^0(z,y)~,\no\\
    &= \tilde G_R^0(p,p') + \int d^4z \[\int d^4x ~e^{ipx}G_R^0(x,z)\] \delta(z^{-}) \(U(\bsz)-1\) 2\p^-_z \[\int d^4y ~e^{-ip'y}G_R^0(z,y)\]~,\no\\
    &= \tilde G_R^0(p,p') - 2i(p')^- \tilde G_R^0(p) \[\int d^4z ~e^{ipz}  \delta(z^{-}) \(U(\bsz)-1\)  e^{-ip'z}\]\tilde G_R^0(p')~,\no\\
    &= \tilde G_R^0(p,p') - 4\pi i(p')^-\delta(p^- -(p')^-)  \tilde G_R^0(p) \[\int d^2\bsz ~e^{i(\bsp-\bsp')\cdot \bsz}\(U(\bsz)-1\) \]\tilde G_R^0(p')~.
\end{align}
In the first line, we isolated the non-interacting piece $G_R^0(p,p') = \tilde{G}_R^0(p) (2\pi)^4\delta^{(4)}(p-p')$ from the rest where $\tilde{G}_R^0(p)$ was defined in Eq.~\eqref{free-G0-prop}. In the second line we distributed the $x$ and $y$ integrals over the nontrivial pieces in the integrand. In the third line, we used the translation invariance of $G_R^0(x,z)$ and $G_R^0(z,y)$ before writing their Fourier transforms $\tilde G_R^0(p)$ and $\tilde G_R^0(p')$ respectively. The derivative $2\p^-_z$ gives the factor of $-2i (p')^-$. In the last line we performed the integral over $z^-$ and $z^+$. The latter integral gives the factor of $2\pi \delta(p^- -(p')^-)$. 

\bibliography{reference}

\end{document}